\documentclass[11pt, singlespacing,nohyperref]{article}
\usepackage[utf8]{inputenc}
\usepackage{authblk}
\usepackage{mathtools}
\usepackage{array}
\usepackage[font=small,labelfont=bf]{caption}
\usepackage{geometry}
\usepackage{graphicx}  
\usepackage[colorlinks=true,citecolor=red,linkcolor =blue]{hyperref}
\usepackage[utf8]{inputenc} 
\usepackage[11pt]{moresize}
\usepackage[T1]{fontenc} 
\usepackage[stable]{footmisc}
\usepackage{changepage}
\usepackage[acronym,nomain]{glossaries}              
\usepackage{tcolorbox}
\tcbuselibrary{theorems}
\newtcbtheorem[number within=chapter]{mytheo}{Subtle Detour}%
{colback=blue!5,colframe=blue!35!black,fonttitle=\bfseries}{th}

\usepackage[backend=bibtex, style=ieee,    dashed=false]{biblatex} 

\usepackage[autostyle=true]{csquotes} 
\usepackage{amssymb}
\usepackage{amsmath}
\newcommand{\normord}[1]{%
  :\mathrel{\mspace{2mu}#1\mspace{2mu}}:%
}

\RequirePackage{geometry}
\title{Electrons Lost in Phase Space}
\author{Tomer Ravid\footnote{NHETC, Rutgers University}}
\date{November 20, 2024}

\begin{document}
\maketitle

\begin{abstract}
I review the formalism of patch bosonization of Fermi surfaces, with a focus on the problem of a two-dimensional metal at a quantum critical point. I argue that this formalism is fundamentally inapplicable to the problem, except in synthetic limits. One such limit is the small-$N$ limit, which was already discussed in early studies of the problem; a similar but slightly less unphysical large-$N$ limit is proposed. I show that it is at least formally possible to construct perturbative expansions around these synthetic limits. However, I argue that nonperturbative effects become important when $N\sim1$.
\end{abstract}

\tableofcontents

\newpage

\section{Introduction}\label{sec:sec1}
Quantum phases of matter are traditionally understood by reducing them to the limit of non-interacting excitations. In the case of fermions, these excitations are Landau quasiparticles. In the case of bosons, these are the density waves of the superfluid. Even exotic, strongly correlated phases such as spin liquids are qualitatively understood in terms of non-interacting fractionalized excitations, which are fluctuations around some mean field solution. This approach does not rest on the assumption that interactions between excitations are \textit{parametrically} weak. Rather, it is sufficient that the interactions do not \textit{qualitatively} change the low-energy physics\textemdash i.e., that no phase transition occurs as one varies the strength of the interaction. This is usually justified on the grounds that at low temperatures few excitations are created, and thus collisions between two excitations are rare\cite{abri}. Consequently, there is a smooth one-to-one mapping from the noninteracting limit to the interacting theory.

However, there are metallic phases that seemingly defy any extrapolation from non-interacting physics. One of the simplest such scenarios\textemdash quite easy to model yet notoriously difficult to analyze\textemdash is that of fermions coupled to a gapless boson in a two-dimensional metal. Such gapless bosons naturally emerge at a ${\boldsymbol{Q}=0}$ quantum critical point, due to the long wavelength fluctuations of the order parameter\footnote{Gapless bosons also arise when there are emergent deconfined gauge fields, which can happen at the transition to a spin liquid or in the half-filled Landau level. The effect of such gauge fields is similar to that of the critical mode, but I will focus on the latter scenario for concreteness.}. The action is

\begin{align}\label{mic}
    S\left[\psi,\bar{\psi},\phi\right]=&\int dt \int d^2p \text{ } \bar{\psi}(\boldsymbol{p},t)\left[i\partial_t-\varepsilon(\boldsymbol{p})\right]\psi(\boldsymbol{p},t) \nonumber\\
    &+\int dt\text{ }d^2 x\text{ } \left[ g\phi(\boldsymbol{x},t)\left|\psi(\boldsymbol{x},t)\right|^2+\dfrac{1}{2}\left(\partial_\mu \phi \right)^2\right],
\end{align}
with $\psi$ the fermion field, $\phi$ the bosonic field and $g$ the coupling constant. The boson's speed has been conveniently set to unity. I am ignoring spin degrees of freedom as well as any non-singular momentum-dependence of the coupling constant\footnote{This means the boson mediates an indiscriminate, isotropic attraction between fermions. Consequently, strictly speaking, the action describes not a symmetry breaking transition, but the onset of phase separation: in the ``ordered'' phase, the fermion fluid contacts under its own attraction into a high-density aggregate, in a way qualitatively analogous to gravitational collapse, and the order parameter is proportional to the density of the aggregate.}, as these details do not affect the universal physics that emerges at the critical point.

The gaplessness of the boson gives rise to singular scattering of electrons near the Fermi surface at zero energetic cost, which turns out to dramatically reorganize the low-energy spectrum. At one loop\cite{Sachdev2}\cite{polchinski}, one finds that the bare fermions are unstable, and have vanishingly small overlap with the actual excitations\textemdash the hallmark of a non-Fermi liquid. Nevertheless, the fermion Green's function at zero frequency retains its pole at the original position of the Fermi surface, a pole which no longer corresponds to a discontinuity in the occupation function $n_{\boldsymbol{k}}$ but to a singularity in its derivative.

Specifically, one finds that the fermions acquire a complex-valued self-energy proportional to $\omega^{2/3}$\textemdash much larger than the bare kinetic term $\omega$ as $\omega\rightarrow 0$. .
Thus, even at vanishingly weak coupling, the effects of the interaction dominate over free fermion behavior and qualitatively change the physics\textemdash the low energy limit fails to commute with the weak coupling limit. Alternatively, it turns that a better way to quantify the strength of the interaction is in terms of the dimensionless coupling-to-mass ratio $\frac{g^2 k_F}{m^2 v_F}$, with $m$ the bosonic mass parameter (equivalently: gap, inverse correlation length) and $v_F$ the Fermi velocity. This ratio can be understood as the magnitude of a point-like potential that produces scattering as strong as does a static boson. It diverges at the critical point regardless of the magnitude of $g$.

The boson-fermion problem is thus physically interesting for the same reason that it is mathematically difficult: it describes electrons that interact so strongly that they seemingly ``forget'' their origin as independent particles. Tackling this problem must consist in finding a limit that is simple enough to be exactly solvable, while retaining enough of the complexity of the strong interactions to be adiabatically connected to the full problem. The task is further complicated by the fact that the fixed point has no adjustable parameters.

One limit that has been widely considered (e.g. \cite{Sachdev2}\cite{polchinski}) is the large-$N$ limit, in which the boson couples to a large number of fermion species. Just like the weak coupling limit, this limit allows one to focus on simple diagrams, and just like weak coupling limit, it doesn't commute with the low energy limit: while each boson interacts with $N$ species of fermions, each fermion only interacts with one species of bosons, so the self-energy of the boson is much greater than the self-energy of the fermion. If the coupling is set so that the former is finite, then the latter must vanish as $N\rightarrow \infty$\textemdash even though for any \textit{finite} $N$ the fermion self-energy dominates over the bare fermion dispersion. If one takes the low-energy limit before the large-$N$ limit, then the smallness of the damping of fermions (which cuts off frequency integrals in the UV) translates to an enhancement of diagrams in the large $N$ limit, and the large-$N$ limit turns out to be uncontrollable\cite{sung-sik-lee}. Recently, Esterlis et al.  proposed a new variant of the large-$N$ theory involving a large number of bosonic flavors and random coupling constants between the various fermionic and bosonic flavors \textit{a-la} SYK\cite{Sachdev}. The self-averaging nature of the random couplings gives rise to a mean-field theory for the propagators and self-energies, and the large-$N$ limit thus exactly reduces to the Eliashberg (i.e., self-consistent one loop) theory. While this is likely the most successful and clever attempt to control the boson-fermion problem to date, it is not established that the crossover from large $N$ to small $N$ is smooth, and that the randomness of the coupling does not give rise to physics qualitatively different from the $N=2$ model (as it certainly does in the SYK model). Finding other exactly solvable cousins of the boson-fermion problem is thus an important task.

One candidate that seems initially promising is that derived from the technique of Fermi surface patch bosonization by Haldane\cite{haldane} and by Neto and Fradkin\cite{neto}, inspired by the early work of Luther\cite{luther}. The basic idea is to focus on electron-hole pairs localized around a point on the Fermi-surface instead of individual electrons. This technique was applied to the boson-fermion problem by several authors \cite{ioffe}\cite{lawler}. Recently, there has been a  surge of interest in bosonization, especially in the context of a proposed extension known as the method of coadjoint orbits\cite{dv}\cite{dv2}\cite{sohn}\cite{mehta}\cite{dv3}. Bosonization seems worthy of consideration for at least two reasons:
\begin{enumerate}
    \item In one-dimensional systems, bosonization is known to account for interactions fully nonperturbatively within a quadratic (thus exactly solvable) model.
    \item In one-dimensional systems, bosonization is the most natural language by which to describe the Luttinger liquid\textemdash the best understood phase of electrons with no fermionic quasiparticles.
\end{enumerate}
Unfortunately, I have become convinced that straightforward bosonization is invalid when applied to the critical metal, except in special synthetic limits that are not perturbatively connected to the realistic model. The literature on Fermi-surface bosonization can be confusing, and there are essential but subtle conceptual issues that are often not clearly addressed. I hope to elucidate these issues while simultaneously providing an intuitive introduction to the main ideas of the formalism. 

\section{Patch Bosonization as a Semiclassical Approximation}
\subsection{Patching Up the Fermi Surface}\label{sec:patch}
In a Fermi gas or a Fermi liquid, fermions close to the Fermi surface disperse linearly\footnote{For notational brevity, I will consider a circular Fermi surface throughout the paper, so that the dispersion depends only on the magnitude of $\boldsymbol{p}$. This will have no effect on the important results.}
\begin{equation}\label{nonloc}
    \epsilon_{\boldsymbol{p}}\approxeq v_F \left(\left|\boldsymbol{p}\right|-k_F\right).
\end{equation}
This seems to be nonlocal when transformed back to position space, which is the main subtlety in formulating effective field theories for Fermi liquids. Such nonlocality is certainly present in the Shankar-Polchinski RG analysis of Fermi liquids\cite{shankar}\cite{polchinski-rg}, which involves scaling momenta closer and closer to the Fermi surface. This nonlocality is unphysical: it is a result of the fact that  ``closeness of momentum to the Fermi surface'' is, by definition, a nonlocal property of the wavefunction. This nonlocality disappears when nonlinear terms in the dispersion are restored (e.g. consider the free fermion dispersion $\frac{k^2}{2m}-\mu$), but those will be irrelevant compared to the linear term at low energies.

Strictly speaking, taking the low energy limit of an ordinary field theory is a nonlocal procedure in a similar sense\textemdash it consists in coarse-graining over a finite length scales. Yet in ordinary field theories, the nonlocality disappears from the action\textemdash it is hiding in the presence of a finite UV cutoff. What makes the nonlocality appear much more explicit in the case of fermions at finite density is the fact that the low energy degrees of freedom of the fermions lie in the vicinity of a \textit{surface} ($\left|\boldsymbol{k}\right|=k_F$) rather than a point ($\boldsymbol{k}=0$) in momentum space. 

This apparent nonlocality in the fermion action is confined to small distances of the order of the Fermi wavelength, which is the minimum uncertainty in position required to specify the shape of the Fermi surface. Thus, it should disappear when one ``zooms out'' to far longer distances.  However, ``zooming out'' in position space is equivalent to ``zooming in'' in momentum space\textemdash and the Fermi surface consists of an infinite number of low-energy points that become increasingly distinct as one zooms in. Thus, the low energy theory must consist of an infinite collection of low-energy fields.  The key to restoring locality is then to linearize each field near its own Fermi point, rather than linearizing a single field near the Fermi \textit{surface} as in eq. \eqref{nonloc}.

\begin{figure}[ht]
    \centering
    \includegraphics[width=0.4\linewidth]{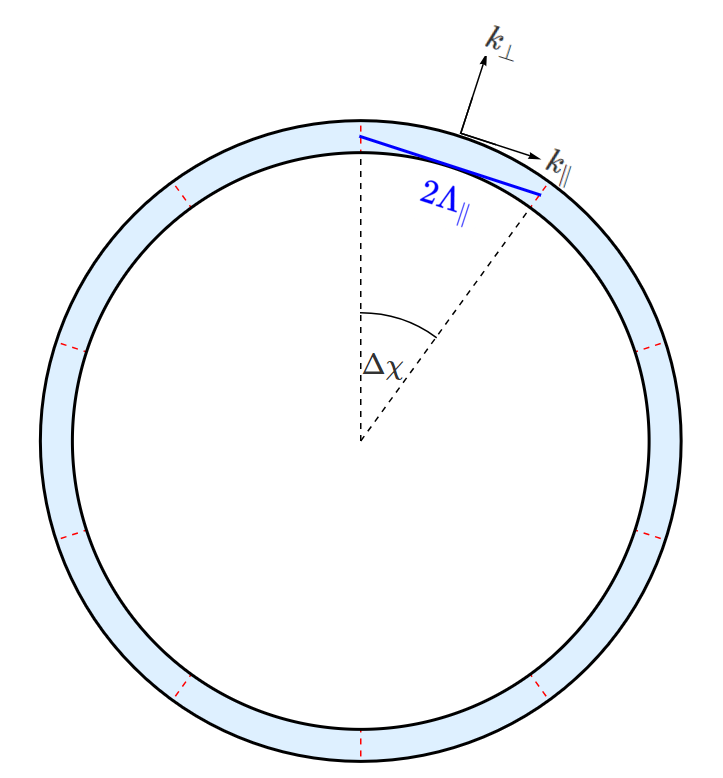}
    \caption{The Fermi surface is broken into patches (indicated by red dashed lines) at angular intervals of $\Delta\chi$. Momenta within the patch are labeled by the normal and tangential coordinates $k_{\perp}$ and $k_{\parallel}$. In order for the patches to fully cover the Fermi surface without overlapping, one must impose a cutoff on $k_{\parallel}$ such that $k_F \Delta\chi=2\Lambda_{\parallel}$.}
    \label{fig:patching}
\end{figure}

Specifically, one can decompose any momentum into a large Fermi momentum and a small deviation from that Fermi momentum:
\begin{align}\label{decomposition}
    \boldsymbol{p}=k_F \boldsymbol{\hat{x}}_{\perp}(\chi)+\boldsymbol{k},
\end{align}
where $\chi$ is an angle around the Fermi surface, and $\boldsymbol{\hat{x}}_{\perp}(\chi)$ is the normal to the Fermi surface, $\boldsymbol{\hat{x}}_{\perp}(\chi)=\cos\chi\boldsymbol{\hat{x}}+\sin \chi \boldsymbol{\hat{y}}$. One can then treat $\boldsymbol{k}$ and $\chi$ as separate variables, and take the Fourier transform with respect to $\boldsymbol{k}$ to obtain a long wavelength, spatially local theory. Of course, if $\chi$ and $\boldsymbol{k}$ are arbitrary, then one is just double counting every total momentum $\boldsymbol{p}$ by treating it as a sum of two independent variables. The decomposition \eqref{decomposition} is unique only if the same momentum $\boldsymbol{p}$ cannot be represented by two different $\chi$s\textemdash i.e., if even the largest possible $\boldsymbol{k}$ is insufficient to take the electron to a different $\chi$. In other words, there must be an upper cutoff on $\boldsymbol{k}$ and a lower cutoff on $\chi$, and the former cannot exceed the latter. Pictorially, this means that the Fermi surface is broken into discrete patches, with $\boldsymbol{k}$ representing a momentum within a patch and $\chi$ labelling the patches (see figure \ref{fig:patching}). The patches fully cover the Fermi surface without overlapping when the cutoff on the component of momentum tangent to the Fermi surface, $k_{\parallel}=k_x\sin\chi-k_y \cos \chi$, coincides with (half) the distance between two neighboring patches, i.e.
\begin{equation}\label{cutoffs}
    \Lambda_{\parallel}=\dfrac{1}{2}k_F \Delta \chi.
\end{equation}

In position space, the decomposition of momenta into a Fermi point $k_F \boldsymbol{\hat{x}}_{\perp}(\chi)$ and a small deviation $\boldsymbol{k}$ translates to a decomposition of the fermion field into a basis of wavepackets, peaked at position $\boldsymbol{y}$ and momentum $k_F \hat{\boldsymbol{x}}_{\perp}(\chi)$:
\begin{align}
    \psi^{\dagger}(\boldsymbol{x})&=\sum_{\chi} \int^{\infty}_{-\infty} \dfrac{dk_{\perp}}{2\pi} \int^{\Lambda}_{-\Lambda} \dfrac{dk_{\parallel}}{2\pi} \exp\left[i \left(k_F\boldsymbol{\hat{x}_{\perp}}+\boldsymbol{k}\right)\cdot\boldsymbol{x}\right] c^{\dagger}_{k_F \boldsymbol{\hat{x}}_\perp+\boldsymbol{k}}
    \nonumber\\
    &=\sum_{\chi} \sum_{\boldsymbol{y}} u\left(\boldsymbol{x},\boldsymbol{y},\chi\right)\psi^{\dagger}(\boldsymbol{y},\chi)
\end{align}
with
\begin{equation}
     \psi^{\dagger}(\boldsymbol{y},\chi)=\int^{\infty}_{-\infty} \dfrac{dk_{\perp}}{2\pi} \int^{\Lambda}_{-\Lambda} \dfrac{dk_{\parallel}}{2\pi} \exp\left[i \left(k_F\boldsymbol{x_{\perp}}+\boldsymbol{k}\right)\cdot\boldsymbol{y}\right] c^{\dagger}_{k_F \boldsymbol{\hat{x}}_\perp+\boldsymbol{k}}
\end{equation}
and
\begin{equation}\label{wavepacket}
u\left(\boldsymbol{x},\boldsymbol{y},\chi\right)=\dfrac{1}{\sqrt{\pi \Lambda}}\delta\left(x_{\perp}-y_{\perp}\right)\exp\left(i k_F y_{\perp}\right) \dfrac{\sin\left(\Lambda {x}_{\parallel}-\Lambda y_{\parallel}\right)}{ x_{\parallel}-y_{\parallel}}.
\end{equation}
and with the sums over $\chi$ and $y_{\parallel}$ appropriately discretized in accordance with \eqref{cutoffs}. By contrast, $x_{\parallel}$ is a continuous variable. (However, note that when $x_{\parallel}$ coincides with the discretization of $y_{\parallel}$ values, $\psi^{\dagger}(\boldsymbol{x})$ is simply a sum of $\psi^{\dagger}(\boldsymbol{x},\chi)$ evaluated at the same spatial point with appropriate phases.) This decomposition can always be performed. What will make it useful is the additional assumption that $y_{\parallel}$ and $\chi$ vary over scales $\delta y_{\parallel}$, $\delta\chi $ much greater than their respective cutoffs, so that the cutoffs can be safely ignored. In other words, the fundamental assumption is that there is some choice of $\Lambda$ for which
\begin{equation}\label{semiclassical}
    \delta y_{\parallel}\gg \dfrac{1}{\Lambda} \gg \dfrac{1}{k_F \delta \chi}.
\end{equation}
This is nothing but a semiclassical approximation: i.e., approximating a given wavepacket \eqref{wavepacket} as infinitely localized in both position and momentum. This is the approximation always employed in the study of Fermi liquids (e.g., to derive the Boltzmann equation)\footnote{Technically, patching still accounts for the uncertainty in $\boldsymbol{x}$ and its conjugate $\boldsymbol{k}$, so long as the latter is much smaller than $k_F \Delta \chi$ in the tangential direction. Thus, the formalism is a generalization of Landau's original semiclassical approximation. This is reflected in the fact that the formalism can be used to directly calculate the quasiparticle lifetime, whereas in Landau's original Hamiltonian implies that it is infinite.} Indeed, it is always justified in a Fermi liquid: in a Fermi liquid, an electron near the Fermi surface scatters off other electrons at any point around the Fermi surface, but only \textit{by} a vanishingly small momentum, so that $\delta\chi\sim 2\pi$ while $\delta y$ is large. It is precisely this approximation that will break down in non-Fermi liquids. 

In a Fermi gas, linearizing the dispersion gives rise to the action of an infinite collection of free noninteracting Weyl fermions
\begin{align}\label{lin}
    S_0\left[\psi,\bar{\psi}\right]&=\int dt \int d^2p \text{ } \bar{\psi}(\boldsymbol{p},t)\left[i\partial_t-\varepsilon(\boldsymbol{p})\right]\psi(\boldsymbol{p},t) \nonumber\\
    &\approxeq \int dt\int_{k_{\parallel\leq \Lambda}} d^2 k \sum_{\chi} \bar{\psi}(t,\boldsymbol{k},\chi) \left[i\partial_t- \left.\dfrac{\partial \varepsilon}{\partial \boldsymbol{k}}\right|_{\boldsymbol{k}=0,\chi}\cdot{\boldsymbol{k}}\right] {\psi}(t,\boldsymbol{k},\chi) \nonumber\\
    &=  \int dt \text{ } dx_{\perp} \sum_{x_{\parallel}}\sum_{\chi} \bar{\psi} (t,\boldsymbol{x},\chi)\left(i\partial_t-i v_F \partial_{\perp} \right)\psi(\boldsymbol{x},t,\chi) \nonumber \\
    &\approxeq  \int dt \text{ } d^2 x \int^{2\pi}_0 d\chi \bar{\psi} (t,\boldsymbol{x},\chi)\left(i\partial_t-i v_F \partial_{\perp} \right)\psi(\boldsymbol{x},t,\chi).
\end{align}
In the second line, the dispersion has been linearized near each patch of the Fermi surface (the slope of the dispersion $\frac{\partial \epsilon}{\partial \boldsymbol{k}}$ at the Fermi surface is the Fermi velocity). In the fourth line, the continuum limit of $\chi$ and $x_{\parallel}$ has been taken, thus ignoring the cutoffs, and $\psi$ has been rescaled by $\sqrt{d\chi}$ so that $\left|\psi(\chi)\right|^2d\chi$ is the density of charge concentrated at patch $\chi$.

In a Fermi liquid, one adds a Landau interaction, which couples the charge density at patch $\chi$ to the charge density at patch $\chi^\prime$
\begin{equation}\label{land-int}
    S_{\text{Landau}}=\int d^2 x dt \int d\chi \int d\chi^\prime f\left(\chi,\chi^\prime\right)\left|\psi\left(\chi,\boldsymbol{x},t\right)\right|^2\text{ } \left|\psi\left(\chi^\prime,\boldsymbol{x},t\right)\right|^2.
\end{equation}
Terms of the form $\psi\left(\chi_1\right)\bar{\psi}\left(\chi_2\right)\psi\left(\chi_3\right)\bar{\psi}\left(\chi_4\right)$, while microscopically legal, are discarded because, by momentum conservation, they necessarily involve scattering far from the Fermi surface, which is energetically penalized. Thus, there is an emergent low-energy conservation of the number of fermions at each patch, and an emergent symmetry under a patch-dependent phase-shift $\psi\rightarrow e^{i\alpha(\chi)} \psi$. This symmetry is an outgrowth of the microscopic symmetry under translation by a lattice vector, which is just one kind of a patch-dependent phase shift. Else et al. conjectured that this symmetry is the essential characteristic of metals even beyond Fermi liquid theory\cite{ersatz}. However, it is not clear in what useful way this symmetry applies in systems whose low energy physics is fundamentally not semiclassical, like the fermion-boson problem in its critical regime, where the low energy physics is essentially local on the Fermi surface (as will be explained in what follows), or like the strange metal where singular large-angle scattering of low-energy fermions is probably responsible for the low temperature $T$-linear resistivity (see \cite{mfl} and \cite{sachdev-strange} for two theoretical models for the strange metal which have no patch-$U(1)$ symmetry and where a patch decomposition would make no sense).

In the critical metal, the interaction mediated by the gapless boson is singular at small momenta. Thus, heuristically, it seems reasonable to expect that a typical boson carries a momentum too small to make an electron hop to a different patch. In other words, it seems reasonable to assume that charge density that couples to the boson in eq. \eqref{mic} can be written as a sum over patches:
\begin{equation}\label{nfl-int}
    S_{g}=g\int dt\text{ }d^2 x \int^{2\pi}_0 d\chi \left|\psi\left(t,\boldsymbol{x},\chi\right)\right|^2 \phi(t,\boldsymbol{x}).
\end{equation}
In Section \ref{sec:trouble}, it will be shown that this is actually not the case, and that the patched theory is inconsistent in the critical regime of the boson-fermion problem. The conventional wisdom attributes this to the linearization of the fermion dispersion. The linearity of the dispersion implies that the patches are flat, and the claim is that flat patches take insufficient account of the curvature of the Fermi surface (see e.g. \cite{chubukov}). However, this is not the precise reason for the failure of the patched theory\textemdash or, at least, it is not the most direct way to state the reason. The flat patch approximation is always exact so long as one takes the continuum limit of patches, as is done in \eqref{nfl-int}. It is the continuum limit itself, when combined with the conservation of patch charge (the absence of inter-patch scattering), that is inconsistent, whether the patches are flat or curved. 

For now, these subtleties will be ignored, and I will discuss the rather reasonable consequences of these unreasonable assumptions.  

\subsection{Bosonized Action}\label{sec:bose-deriv}
Whether one considers the Landau interaction \eqref{land-int} or the patched critical interaction \eqref{nfl-int}, the linearized action \eqref{lin} describes an infinite collection of quasi-one dimensional Weyl fermions dispersing in different directions, and with no mixing between different flavors. Thus, it is exactly solvable for the same reason that the interacting one-dimensional electron gas is exactly solvable. By analogy with the one-dimensional case, one can bosonize each of the Weyl fermions by treating the normal position $x_{\perp}$ as the only spatial direction, and the tangential position $x_{\parallel}$ and patch angle $\chi$ as flavor labels. The result is a bosonic action for an infinite collection of linearly dispersing electron-hole fields.

\begin{figure}
    \centering
    \includegraphics[width=0.4\linewidth]{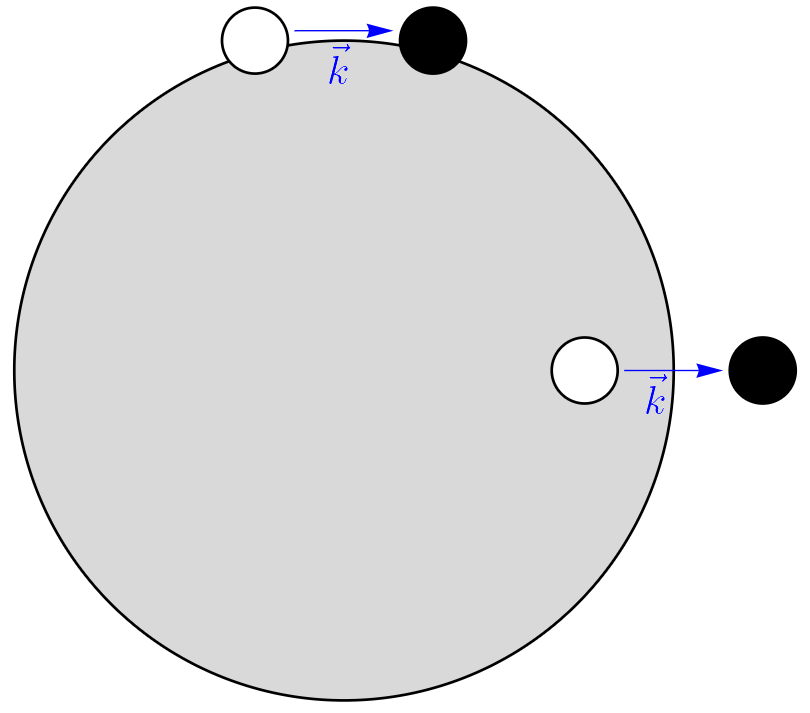}
    \caption{An electron-hole pair of momentum $\boldsymbol{k}$ can have a continuum of energies: if $\boldsymbol{k}$ is locally tangential to the Fermi surface, then the pair's energy vanishes, whereas if $\boldsymbol{k}$ is locally normal to the Fermi surface, then the energy is $v_F k$.}
    \label{fig:continuum}
\end{figure}

The continuum of bosonic fields is nothing but the standard Lindhard continuum, and its physical origin is intuitive. An electron-hole pair of a \textit{given net momentum} $\boldsymbol{k}$ can have a continuum of energies depending on its location along the Fermi surface (essentially, the relative momentum of the pair): if $\boldsymbol{k}$ is locally tangent to the Fermi surface, then the energy of the pair will vanish (since the electron and the hole are equally far from the Fermi surface); if $\boldsymbol{k}$ is locally normal to the Fermi surface, the energy will be given by $v_F\left|\boldsymbol{k}\right|$; for any other point along the Fermi surface, the energy can take any value between these two extremes, and is given by $v_F k_{\perp}$ (see figure \ref{fig:continuum}). This input is sufficient to ``guess'' the correct form of the Gaussian action of the free Fermi gas in terms of the bosonized fields $\zeta(t,\boldsymbol{x},\chi)$:

\begin{equation}
    S_\zeta =A \int dt\text{ }d^2 x \int^{2\pi}_0 d\chi \partial_{\perp} \zeta \left(\partial_{t} \zeta - v_F\partial_{\perp} \zeta  \right).
\end{equation}
This is a chiral bosonized action, in the sense that all Weyl fermions have been bosonized as one-dimensional ``right movers.'' Chiral bosoniztaion will prove more convenient once non-Gaussian corrections are included.

The coefficient $A$ can be determined by a systematic derivation, but there is a ``quick and dirty'' way to obtain its value, by analogy with the one-dimensional case: The consistency relation between the cutoffs \eqref{cutoffs} implies that $x_{\parallel}$ and $\chi$ must be discretized so that
\begin{equation}\label{uncert}
    \Delta x_{\parallel} \Delta \chi=\dfrac{2\pi}{k_F}.
\end{equation}
Plugging this into the action in place of the differentials $ dx_{\parallel} d\chi$ gives
\begin{equation}
    S_\zeta=\dfrac{2\pi A}{k_F} \int dt\text{ }dx_{\perp} \sum_{m,n} \left[\partial_{t} \zeta_{m,n} \partial_{\perp} \zeta_{m,n}- v_F \left(\partial_{\perp} \zeta_{m,n}\right)^2 \right],
\end{equation}
with $x_{\parallel}=\Delta x_{\parallel} n$ and $\chi=\Delta \chi m$. The action now takes the form of a \textit{discrete} set of decoupled one-dimensional leads. Recall that in the theory of one-dimensional Luttinger liquids, the coefficient $A$ determines the exponent of the fermion propagator:
\begin{equation}
    \left<\bar{\psi}_{m,n} \left(x_{\perp},t\right) \psi_{m,n}(0,0)\right>\sim\dfrac{1}{\left(x_{\perp}-v_F t\right)^{\frac{8\pi^2 A}{k_F}}},
\end{equation}
so the correct propagator for a free Fermi gas is obtained when 
\begin{equation}
    A=\frac{k_F}{8\pi^2}.
\end{equation}
This is a first example of the necessity of imposing the correct UV cutoffs in order to obtain sensible IR physics, a recurring theme throughout this paper. 

Interactions like \eqref{land-int} and \eqref{nfl-int} can be added by noting that just like in one dimension, the charge density (per patch angle) should be proportional to the spatial derivative of the electron-hole field
\begin{equation}
    \rho=\dfrac{k_F}{4 \pi^2} \partial_{\perp} \zeta,
\end{equation}
where the prefactor can, again, be verified using the cutoff consistency relation \eqref{uncert}. The fact that the charge density per-patch is linear in $\zeta$ means that a patch-conserving two-body interaction will be quadratic in $\zeta$, leading to exact solubility. With either the Fermi liquid or critical interaction, the equation of motion for $\rho$ will take a form similar to a collisionless Boltzmann equation in the linear response regime. 

These simple arguments can be translated into a systematic derivation, for example, in the path integral picture. The derivation is completely analogous to the one-dimensional case (e.g. \cite{chir}). Here, I just outline the main steps, focusing on the critical metal:
\begin{enumerate}
    \item Start with the action $S=S_0+S_g+S_{\phi}$, with $S_0$ and $S_g$ given in \eqref{lin} and \eqref{nfl-int}, and $S_{\phi}$ the bosonic part of the action \eqref{mic}. 
    \item Since $\phi(\boldsymbol{x},t)$ couples to the charge density, it acts on the quasi-one-dimensional Weyl fermions as if it were an electric potential, and its effect is to shift their phase. It turns out that by locally shifting the phase of $\psi$ by some appropriate patch-dependent $\tilde{\zeta}(t,\boldsymbol{x},\chi)$, one can eliminate the interaction of $\psi$ with $\phi$. Use this to integrate out the phase-shifted fermions.
    \item However, the phase shift doesn't come for free: due to the one-dimensional chiral anomaly, the patch-dependent phase shift changes the measure of the path integration (see \cite{fujikawa} for a derivation applicable to the analogous 1+1-dimensional case). The transformation of the measure turns out to be a phase factor quadratic in $\tilde{\zeta}$, which can be recast as a Gaussian path integral over a quantized $\zeta$. The resulting action is precisely the bosonized action, given by 
    \begin{align}\label{bosonized-action}
    S[\zeta,\phi]=&\dfrac{k_F}{8\pi^2} \int^{2\pi}_{0} d\chi  \int dt\text{ } d^2x \partial_{\perp} \zeta\left(\partial_{t} \zeta- v_F \partial_{\perp} \zeta+2g \phi \right)\nonumber\\
    & +\dfrac{1}{2}\int dt\text{ }d^2 x \left[\left(\partial_\mu \phi\right)^2+m_0^2 \phi^2\right],
\end{align}
\end{enumerate}
where a bare bosonic mass has been introduced because the mass will turn out to be renormalized. This derivation establishes that $\zeta$ is a $2\pi$-periodic phase, much like in the one-dimensinoal case. It can be extended to any correlators by inserting fermion fields into the path integral and repeating the argument. This leads to a bosonization formula relating correlators of $\psi$ to correlators of the phase shift $e^{i\zeta}$. 

The chiral anomaly of the patch $U(1)$ symmetry invoked in the derivation has been argued by Else et al. to be another essential characteristic of metals\cite{ersatz}. The authors used it to prove the Luttinger theorem without explicitly assuming a Fermi liquid. In a subsequent paper, this anomaly was applied to the boson-fermion problem to determine the exact boson propagator at zero momentum, and to argue that the optical conductivity is given by a free Drude peak\cite{goldman}. These last results can be easily obtained from the bosonized theory.

\section{Properties of the Bosonized Solution}\label{sec:bose-res}
\subsection{Boson Propagator}\label{sec:bose-prop}
\begin{figure}
    \centering
    \includegraphics[width=0.25\linewidth]{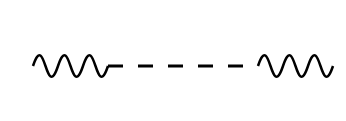}
    \caption{The only boson self-energy diagram in the bosonized theory. Wavy lines are bare bosonic propagators, while dashed lines are bare electron-hole propagators.}
    \label{fig:boson-prop}
\end{figure}

Since the bosonized action \eqref{bosonized-action} is quadratic, the only thing a boson can do is turn into an electron-hole pair of the same energy and momentum but of any patch, and the only thing this electron-hole pair can do is turn back into a boson (see figure \ref{fig:boson-prop}). Thus the boson self-energy consists on a single electron-hole propagator, integrated over all patches but not over energy or momentum:
\begin{align}\label{boson-se}
    \Pi(\omega,\boldsymbol{k})&=\dfrac{g^2 k_F}{4\pi^2} \int^{2\pi}_0 d\chi \dfrac{k^2_{\perp}}{k_{\perp}\left(\omega-v_F k_{\perp}+i\epsilon \right)}\nonumber\\
    &=\dfrac{g^2 k_F}{2\pi v_F}\left(-1+i\dfrac{\left|\omega\right|}{\sqrt{v_F^2 k^2-\omega^2}}\right).
\end{align}
The first term is a \textit{negative} shift of the bosonic mass. It can be understood as a ``reverse screening'' of the \textit{attractive} interaction mediated by the boson: every electron surrounds itself with a cloud of other electrons, in turn allowing it to attract even farther electrons and extending the effective range of the interaction. I set the positive bare mass $m_0^2$ in \eqref{bosonized-action} to cancel this shift, so as to tune the metal to criticality.

The second term is just the famous density fluctuation spectrum of a two-dimensional Fermi gas. A similar term famously arises in the renormalization of the electrostatic potential in a metal within the Lindhard theory, except that its low-energy effect on electrons is not as singular even in two dimensions since the electrostatic interaction is screened. The term is purely imaginary when $\omega<v_F k$ (i.e., within the electron-hole continuum) and purely real when $\omega>v_F k$ (i.e., above the electron-hole continuum). 

 When $\omega<v_F k$, the imaginary contribution corresponds to a decay of bosonic excitations into the continuum (Landau damping), eliminating the poles of the propagator except at $\omega=k=0$. The contribution to the self-energy in this regime is purely due to the residue from the poles of the integrand in eq. \eqref{boson-se}. These poles come from a pair of antipodal points around the Fermi surface resonant with the boson, i.e., at which the bare dispersion $\omega=v_F k_{\perp}$ is obeyed by the boson's energy and momentum. Since there is a \textit{continuum} of electron-hole states, the boson can \textit{always} find such a pair of points. The precise dependence on $\omega$ and $\boldsymbol{k}$ can be interpreted as the number of electron-hole excitations to which a bosonic wavepacket with some small \textit{range} of frequencies $\delta \omega$ can decay while satisfying the resonance condition $\omega=v_F k_{\perp}$.

When $\omega>v_F k$, the real contribution is a shift of the pole of the boson's propagator and thus a renormalization of the dispersion of bosonic excitations. However, due to the bare mass required to cancel the ``reverse screening,'' these bosonic excitations become gapped, as though repelled by the boson-thirsty electron-hole continuum. Indeed, the renormalization of the boson's dispersion ensures that the bosonic excitations can never intersect the continuum regardless of the bare speed of bosons.

The most striking thing about the result \eqref{boson-se} is that at low energies and momenta, the self-energy  \textit{always} dominates over the bare dynamics of the bosons, since 
\begin{equation}
    \omega^2\ll\dfrac{g^2 k_F}{2\pi v_F} \dfrac{\left|\omega\right|}{\sqrt{v_F^2 k^2-\omega^2}}.
\end{equation}
At the critical point, the boson's propagator is then peaked near momenta
\begin{equation}
    k^2\sim \dfrac{g^2 k_F}{2\pi v_F}\dfrac{\left|\omega\right|}{\sqrt{v_F^2 k^2-\omega^2}}\gg \omega^2,
\end{equation}
and quantities like the fermion Green's function will be dominated by such bosonic modes. Thus, low energy electrons predominantly scatter off bosons deep inside the electron-hole continuum $\omega\ll v_F k$. For such bosons, one can expand \eqref{boson-se} to leading order in $\omega$ and obtain a propagator
\begin{equation}\label{boson-prop}
    D(\omega,\boldsymbol{k})=\dfrac{1}{k^2+i \dfrac{g^2 k_F}{2\pi v_F^2}\dfrac{\left|\omega\right|}{\left|\boldsymbol{k}\right|}},
\end{equation}
with the $\omega^2$-term discarded as irrelevant. This bosonic propagator has a symmetry under a scaling of space and time with dynamical exponent $z=3$, i.e. $t$ scales as $r^3$. In fact, it will be shown that it gives rise to a critical regime in which the fermions obey similar scaling.

\subsection{Electron-Hole Propagator and Renormalized Electron-Hole Continuum}\label{sec:eh-continuum}
\begin{figure}
    \centering
    \includegraphics[width=0.5\linewidth]{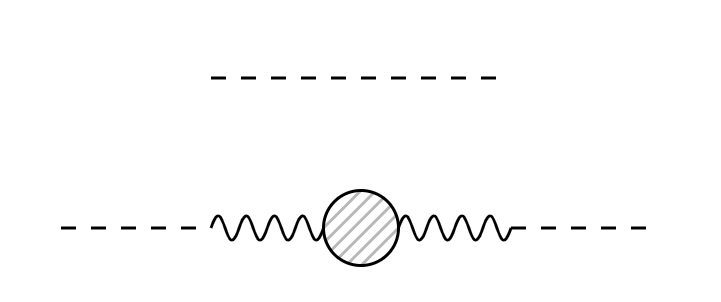}
    \caption{The only two diagrams contributing to the electron-hole propagator. A dashed line is a bare electron-hole propagator, while a wavy line with a bubble is a full boson propagator.}
    \label{fig:eh-se}
\end{figure}

Similarly to the above discussion, the only thing an electron-hole pair can do is turn into a boson, which can only turn back into an electron-hole pair. However, the final electron-hole pair can occupy a patch different from that of the initial electron-hole pair\textemdash a scenario for which the bare propagator does not account. The electron-hole pair can either not scatter into a boson, in which case its patch is conserved, or scatter into a boson, in which case its patch definitely changes (see figure \ref{fig:eh-se}). The full bosonized electron-hole propagator is thus a sum of the patch-local bare propagator and the patch-uncorrelated $t$-matrix term, the latter of which is a product of two external bare electron-hole propagators and one dressed boson propagator:
\begin{align}\label{eh-prop}
        \Delta \left(\omega,\boldsymbol{k},\chi_1,\chi_2\right)=&\dfrac{4\pi^2}{k_F}\dfrac{1}{k_{\perp}\left(\omega-v_F k_{\perp}\right)}\delta\left(\chi_1-\chi_2\right) \nonumber\\
        &+g^2\dfrac{1}{\omega-v_F k_{\perp}\left(\chi_1\right)} \dfrac{1}{k^2+i\dfrac{g^2 k_F}{2 \pi v^2_F}\dfrac{\left|\omega\right|}{\left|\boldsymbol{k}\right|}} \dfrac{1}{\omega-v_F k_{\perp}\left(\chi_2\right)}.
\end{align}
Unlike the $\omega<v_F k$ boson propagator, the electron-hole propagator has retained its pole; in fact, it seems to have acquired an \textit{additional} pole. All the poles occur at $\omega<v_F k$ as in the free theory. This suggests that there is stable continuum of particle-hole excitations, despite the absence of long-lived single-particle excitations in a non-Fermi liquid. 

Indeed, owing to its quadraticity, the action \eqref{bosonized-action} can be easily diagonalized. However, since this chiral action  is linear in time derivatives, the different $\zeta$s are not actually independent degrees of freedom, and they do not commute at different spatial points. One can eliminate the redundant degrees of freedom by combining antipodal patches $\chi$ and $\chi+\pi$ (which naturally move in opposite directions) into a single field, much like one combines  right movers and left movers in the one-dimensional case, defining:
\begin{align}
     \zeta_{\pm}(\chi)=\dfrac{\zeta(\chi)\pm \zeta(\chi+\pi) }{2}
\end{align}
where $\chi$ now runs from $0$ to $\pi$ to avoid double-counting. Plugging this into the action, one finds that only $\zeta_{-}$ couples to $\phi$. By integrating $\zeta_{+}$ out one obtains the action of coupled harmonic oscillators:
\begin{align}\label{osc-action}
    S[\zeta,\phi]=\dfrac{k_F }{16\pi^2 v_F} \int^{\pi}_{0} d\chi \int dt\text{ } d^2r&\left[\left(\partial_{t} \zeta_{-}\left(t,\boldsymbol{r},\chi\right)\right)^2-v_F^2\left(\partial_{\perp} \zeta_{-}\left(t,\boldsymbol{r},\chi\right)\right)^2\right.\nonumber\\
    &\left.+2g v_F\phi(t,\boldsymbol{r})\partial_{\perp}\zeta_{-}\left(t,\boldsymbol{r},\chi\right)\right]+S_{b}[\phi].
\end{align}
The oscillators can be decoupled by writing them as superposition of normal modes:
\begin{align}\label{normal}
    \zeta_{-}(t,\boldsymbol{k},\chi)&=\sum_{\lambda} a_{\lambda}(t,\boldsymbol{k}) u_{\lambda} (\boldsymbol{k},\chi) \nonumber \\
    \phi(t,\boldsymbol{k})&=\sum_{\lambda} a_{\lambda}(t,\boldsymbol{k}) v_{\lambda} (\boldsymbol{k})
\end{align}
where $\lambda$ is the energy of the mode, $a_{\lambda}(t,\boldsymbol{k})$ are some coefficients, and $u_{\lambda} (\boldsymbol{k},\chi)$, $v_{\lambda} (\boldsymbol{k})$ are the solutions to the eigenvalue equations
\begin{align}\label{eigen}
    \nonumber \left(\lambda^2-v_F^2 k_{\perp}^2\right)u_{\lambda} (\boldsymbol{k},\chi)&=- g v_F k_{\perp} v_{\lambda}(\boldsymbol{k}) \\
    \left(\lambda^2-k^2-m_b^2\right) v_{\lambda} (\boldsymbol{k})&=\dfrac{g k_F}{4\pi^2} \int^{\pi}_0 d\chi k_{\perp}  u_{\lambda} (\boldsymbol{k},\chi),
\end{align}
which are just the classical equations of motion for modes of frequency $\lambda$.

The solutions of \eqref{eigen} change their character depending, again, on whether $\lambda$ is inside the electron-hole continuum ($\lambda<v_F k$) or above the electron-hole continuum ($\lambda>v_F k$)\textemdash i.e., depending on whether the left hand side of the top equation in \eqref{eigen} vanishes for some choice of $\chi$, thereby rendering this equation non-invertible. In the latter case, one finds a gapped branch which is nothing but the pole in the bosonic propagator obtained in Section \ref{sec:bose-prop}. In the former case, one can always define a ``renormalized patch'' $\chi^{\star}$ so that $\lambda=v_F \boldsymbol{k}\cdot\boldsymbol{\hat{x}}_{\perp}\left(\chi^{\star}\right)$. It turns out that there is a continuum of excitations for all choices of $\chi^{\star}$. Since this continuum is clearly analogous to the free electron-hole continuum, it is convenient to rewrite \eqref{normal} in terms of $\chi^{\star}$ instead of $\lambda$ and denote the coefficients by $\tilde{\zeta}$
\begin{align}\label{normal-2}
    \zeta_{-}(t,\boldsymbol{k},\chi)&=\int^{\pi}_{0} d\chi^{\star} \tilde{\zeta} \left(t,\boldsymbol{k},\chi^\star\right) u \left(\boldsymbol{k},\chi,\chi^{\star}\right) \nonumber \\
    \phi(t,\boldsymbol{k})&=\int^{\pi}_{0} d\chi^{\star} \tilde{\zeta} \left(t,\boldsymbol{k},\chi^\star\right) v \left(\boldsymbol{k},\chi^{\star}\right).
\end{align}
To simplify the notation, assume that $\boldsymbol{k}$ is in the $x$-direction, so that $\boldsymbol{k}\cdot\boldsymbol{\hat{x}}_{\perp}(\chi)=k\cos\chi$ and likewise for $\chi^{\star}$. Then up to sub-leading terms, the solution to the normal modes equations at the critical point turns out to be
\begin{align}
    \tilde{\zeta}\left(\chi^{\ast},\boldsymbol{k},t\right)=& \sin \chi^{\star} \mathcal{P}\int^{\pi}_0 d\chi \dfrac{\cos \chi \cos \chi^{\star}}{\cos^2 \chi^{\star}-\cos^2\chi}\zeta_{-}(\chi,\boldsymbol{k},t)
\end{align}
where $\mathcal{P}$ denotes principal value\textemdash i.e., discarding the residue from the pole at $\chi=\chi^{\star}$. The residue from the pole has disappeared at the lowest energies after normalizing the normal modes. Thus, the normal mode has zero overlap with the bare excitation at $\chi=\chi^{\star}$! The solution does not depend on $g$ (the dependence on $g$ is in the subleading terms), and there is clearly no adiabatic continuity of the excitations as the interaction is ``turned on.'' Thus, the low-lying excitations are electron-hole pairs that scatter so strongly around the Fermi surface that they lose their connection to the bare electron-hole pair, even at vanishingly small $g$. By solving the normal mode equations away from criticality, it can be also shown that, since $\zeta$ is a $2\pi$ periodic variable, $\tilde{\zeta}$ has a periodicity that becomes larger and larger as one approaches the critical point, and so at the critical point it cannot be re-fermionized by taking the exponential of $\tilde{\zeta}$.

However, it should not be too surprising that that the bosonized action admits two-body quasiparticles. Indeed, any quadratic action admits quasiparticles that are linearly related to the original fields. It is an interesting question whether the continuum of electron-hole excitations is a mere artifact of the quadraticity of the bosonized action, or whether it is a true feature of the quantum critical metal. In the non-bosonized theory, one may attempt to estimate the ``electron-hole propagator'' using RPA fermion propagators\textemdash this does not reveal any trace of continuum of poles. Guo\cite{guo} studied the fluctuation spectrum of fermion bilinears (equivalent to electron-hole modes) more systematically within the random coupling large $N$ theory. He found that, like the fermions field, their dynamics is highly nonlocal in time, which rules out their interpretation as excitations. It is likely that the true excitations of the critical metal are far more exotic and than the  electron-hole pairs of the bosonized theory, and that they involve a macroscopic number of electron-hole pairs.

\subsection{Fermion Propagator}\label{sec:ferm-prop}
By analogy with one-dimensional bosonization, or by the arguments sketched in Section \ref{sec:bose-deriv}, the fermion propagator is given by 
\begin{align}\label{prop}
    \nonumber \left<\bar{\psi}\left(\boldsymbol{x}^\prime,t^\prime,\chi\right) {\psi}\left(0,0,\chi^\prime\right)\right>&=\left<\exp\left[i\zeta \left(\boldsymbol{x},t,\chi\right)\right] \exp\left[-i\zeta\left(\boldsymbol{0},0,\chi^\prime\right)\right]\right>\\ \nonumber
    &=\exp\left[\left<\zeta\left(\boldsymbol{x},t,\chi\right)\zeta \left(\boldsymbol{0},0,\chi^\prime\right)\right>\right],
\end{align}
where the rules of Gaussian integration have been used in the last equality. From \eqref{eh-prop}, the full electron-hole propagator is
\begin{equation}
     \Delta\left(\boldsymbol{x},t,\chi,\chi^\prime\right)=\Delta_{0}\left(\boldsymbol{x},t,\chi,\chi^\prime\right)+\Delta^{\prime}\left(\boldsymbol{x},t,\chi,\chi^\prime\right),
\end{equation}
where $\Delta_0$ is the free electron-hole propagator and $\Delta^\prime$ is the $t$-matrix term. Thus, the fermion propagator must be the product of the free fermion propagator and a scattering correction:
\begin{align}\label{label}
     G\left(\boldsymbol{x},t,\chi,\chi^\prime\right)&= G_0 (\boldsymbol{x},t) G^{\prime} \left(\boldsymbol{x},t,\chi,\chi^\prime\right)\nonumber\\
     &=\dfrac{\delta\left(x_{\parallel}\right) \delta\left(\chi-\chi^\prime\right)}{x_{\perp}-v_F t} e^{\Delta^{\prime}\left(\boldsymbol{x},t,\chi,\chi\right)}.
\end{align}
The task is then to compute the Fourier transform of the $t$-matrix term evaluated at $\chi=\chi^\prime$ and $x_{\parallel}=0$. At low energies one can safely assume that the bosonic momentum is nearly tangential to the Fermi surface, $k=k_{\parallel}$, since tangential electron-hole pairs cost zero energy. Thus 
\begin{equation}
    \Delta^\prime(\omega,\boldsymbol{k},\chi)=g^2\dfrac{1}{\left(\omega-v_F k_{\perp}\right)^2} \dfrac{1}{k_{\parallel}^2+i\dfrac{g^2 k_F}{2 \pi v_F^2}\dfrac{\left|\omega\right|}{\left|k_{\parallel}\right|}}.
\end{equation}
This is the product of a function of $\omega-v_F k_{\perp}$ and a function $\omega$ and $k_{\parallel}$; hence, its Fourier transform will be a product of a function of $x_{\perp}$ and a function of $x_{\perp}-v_F t$ and $x_{\parallel}$($=0$). The $k_{\perp}$ Fourier transform gives
\begin{equation}\label{class-m}
    \int^{\infty}_{-\infty} \dfrac{dk_{\perp}}{2\pi}  \dfrac{e^{-i k_{\perp} x_{\perp}}}{\left(\omega-v_F k_{\perp}\right)^2}=-\dfrac{1}{2 v_F^2} \left|x_{\perp}\right| \exp\left(\dfrac{i}{v_F}\omega x_{\perp}\right).
\end{equation}
The $k_{\parallel}$ Fourier transform evaluated at $x_{\parallel}=0$ is just an integral that can be computed by dimensional analysis
\begin{align}
    \int^{\infty}_{-\infty} \dfrac{dk_{\parallel}}{2\pi}\dfrac{1}{k_{\parallel}^2+i\dfrac{g^2 k_F}{2\pi v_F^2} \dfrac{\left|\omega\right|}{\left|k_{\parallel}\right|}}&=\left(\dfrac{g^2 k_F}{2 \pi v_F^2}\right)^{-1/3}\left|\omega\right|^{-1/3} \int^{\infty}_{-\infty} \dfrac{ds}{2\pi}\dfrac{1}{s^2+\frac{i}{\left|s\right|}}\nonumber\\
    &=\dfrac{2}{3\sqrt{3}} \left(\dfrac{g^2 k_F}{2\pi v_F^2}\right)^{-1/3}  e^{-i\frac{\pi}{6} }\left|\omega\right|^{-1/3}.
\end{align}
Finally,
\begin{align}\label{finally}
\nonumber    \Delta^{\prime}\left(x_{\perp},x_{\parallel}=0,t\right)&=-\dfrac{2}{3\sqrt{3}} \left(\dfrac{2\pi v_F^2 g^4}{k_F}\right)^{1/3}  \left|x_{\perp}\right| \int^{\infty}_{-\infty} e^{-i\frac{\pi}{6}} \dfrac{d\omega}{2\pi}  \left|\omega\right|^{-1/3} \exp\left[\dfrac{i}{v_F}\omega \left(x_{\perp}-v_F t\right)\right]\\
&= -\dfrac{2}{3\sqrt{3}} \Gamma\left(\dfrac{2}{3}\right)\left(\dfrac{v_F^2 g^4}{4\pi^2k_F}\right)^{1/3}  e^{-i\frac{\pi}{6}}\dfrac{\left|x_{\perp}\right|}{\left(x_{\perp}-v_F t\right)^{2/3}},
\end{align}
and
\begin{equation}\label{ferm-prop}
    G(\boldsymbol{x},t)=\dfrac{\delta\left(x_{\parallel}\right)}{x_{\perp}-v_F t}\exp\left[-\kappa e^{-i\pi/6}\dfrac{\left|x_{\perp}\right|}{\left(x_{\perp}-v_F t\right)^{2/3}}\right],
\end{equation}
with $\kappa$ the complicated prefactor from \eqref{finally}. This result was originally obtained in \cite{ioffe}. It is consistent with the equal time result of \cite{lawler} but inconsistent with their equal position result. 

The propagator \eqref{ferm-prop} has the form reminiscent of anomalous diffusion. It describes a zero that moves at the Fermi velocity and becomes broader and broader as it moves. It is invariant under the scaling
\begin{align}
    x_{\perp}&\rightarrow \alpha x_{\perp} \nonumber\\
    x_{\perp}-v_F t&\rightarrow \alpha^{3/2} \left(x_{\perp}-v_F t\right) \nonumber \\
    \psi(\boldsymbol{x},t)&\rightarrow \alpha^{-3/4} \psi(\boldsymbol{x},t).
\end{align}
This scaling is consistent with what one obtains from the one-loop fermion propagator with the $\omega^{2/3}$ self-energy; however, the functional \textit{form} is quite different. This applies, in particular, to one-loop calculations performed within the theory of patched fermions, which is equivalent to the bosonized theory. One dramatic new feature of \eqref{ferm-prop} is that its Fourier transform \textit{vanishes} at zero frequency and nonzero momentum, whereas the one loop propagator is simply given by $\frac{1}{\epsilon_{\boldsymbol{k}}}$ at zero frequency. The conclusion is that, while the boson self-energy calculated within bosonization is exactly given by the one-loop polarization bubble, the fermion propagator takes into account an infinite number of diagrams with an arbitrary number of loops. This creates the hope that bosonization is a way to control the strong interactions beyond perturbation theory\textemdash a hope that will soon be dispelled.

\subsection{Scale Invariance}\label{sec:scaling}
The scaling behavior of the fermion propagator \eqref{ferm-prop} is a manifestation of a consistent low-energy symmetry of the theory. To determine the full symmetry, consider the real-space electron-hole propagator evaluated near $x_{\parallel}=0$ and for nearly equal patches, as was essentially done in Section \ref{sec:ferm-prop}:

\begin{align}\label{scaling-prop}
\Delta\left(x_{\perp},x_{\parallel}\approxeq 0,t,\chi\approxeq\chi^\prime\right)=&\delta(\chi-\chi^\prime)\delta\left(x_{\parallel}\right)\log \left(x_{\perp}-v_F t\right)\nonumber\\
&-\kappa e^{-i\frac{\pi}{6}}\dfrac{\left|x_{\perp}\right|}{\left(x_{\perp}-v_F t\right)^{2/3}}.
\end{align}
Under what scaling is the form of this propagator preserved? Since $\zeta$ is a $2\pi$-periodic variable, it cannot scale, and the electron-hole propagator must remain \textit{invariant}. Invariance of the first term (treating the logarithm as dimensionless) then necessitates that if one scales $x_{\parallel}$ to infinity
\begin{equation}\label{y-scaling}
    x_{\parallel}\rightarrow s^{-1} x_{\parallel}
\end{equation}
one must also scale the patch angle to zero
\begin{equation}
    \chi\rightarrow s\chi.
\end{equation}
In other words: one is zooming closer and closer to a specific point around the Fermi surface\footnote{Of course, there is nothing special about $\chi=0$, and one could just as well scale into any other patch, consistent with the fact that the delta function in eq. \eqref{scaling-prop} only depends on $\chi-\chi^\prime$.} \footnote{In principle, one should also account for patches near the antipodal point $\chi=\pi$, which are equally resonant with the boson and equally relevant under scaling. The antipodal patch is important for many results (for example, it modifies the critical exponents, compare \cite{chiral} to \cite{Sachdev2}), but it will not affect the results of this paper. I will thus ignore the antipodal patches in the interest of notational brevity.}. This is a unique feature of the boson-fermion criticality. It is a result of the tendency of a boson with a given momentum to resonate with a specific point around the Fermi surface as discussed in \ref{sec:bose-prop}\textemdash at low energies, the resonant point will be such that the bosonic momentum is nearly tangential to the Fermi surface, for the reason explained below eq. \eqref{label}. Hence, degrees of freedom at different patches and with different directions of the bosonic momentum decouple.

Due to the scaling of the patch angle to zero, $x_{\parallel}$ can be expanded as
\begin{equation}
    x_{\parallel}=y \cos\chi-x\sin\chi= y-x\chi +\mathcal{O}\left(\chi^2\right),
\end{equation}
The first two terms can be made covariant under the scaling \eqref{y-scaling} if
\begin{align}
    y&\rightarrow s^{-1} y \nonumber\\
    x&\rightarrow s^{-2} x.
\end{align}
Hence, the scaling is anisotropic, which is perfectly acceptable since scaling close to a specific patch breaks rotational symmetry in the first place. The fact that $x$ scales as the \textit{square} of $y$ implies that the scaled Fermi surface looks like a parabola\textemdash as any curve locally does. The dispersion in the patched theory turns into
\begin{equation}
    \epsilon(\boldsymbol{k})=v_F k_x+v_F k_y\chi +\text{irrelevant terms},
\end{equation}
and in the unpatched theory takes the form
\begin{equation}
    \epsilon\left(\boldsymbol{p}\right)=v_F p_x+ \dfrac{v_F}{2k_F} p_y^2+\text{irrelevant terms},
\end{equation}
the latter of which is the standard ``single-patch dispersion'' (though throughout this paper I use the word ``patch'' in a more specific sense, so that an ``unpatched single-patch theory'' is not a contradiction). This dispersion is the starting point for many standard treatments of the boson-fermion problem, for precisely the same reason. 

For $x_{\perp}$, one finds
\begin{equation}
    x_{\perp}=x+\text{subleading terms},
\end{equation}
and by plugging this into the second term of \eqref{scaling-prop}, one finds the result is \textit{almost} invariant if 
\begin{equation}
    t\rightarrow s^{-3} t
\end{equation}
except that the $x_{\perp}$ term in the denominator will then scale to zero, which simply means that it is irrelevant. This scaling is consistent with the fermion propagator, and gives  $\psi\left(t,\boldsymbol{x},\chi\right)\rightarrow s^{2} \psi\left(t,\boldsymbol{x},\chi\right)$. It is also consistent with the boson propagator, and gives  $\phi\left(t,\boldsymbol{x},\chi\right)\rightarrow s^2 \phi\left(t,\boldsymbol{x},\chi\right)$.

This scaling, which is a completely standard result  (aside from the scaling of $\chi$; see e.g. \cite{Sachdev2}), is already embodied in the bare patched action:
\begin{equation}
    S=\int dt\text{ }d^2x\left[\int^{2\pi}_0 d\chi \bar{\psi}\left( i\partial_{t} -i v_F \cos\chi \partial_x-i v_F\sin\chi\partial_{y} +g\phi \right)\psi+\dfrac{1}{2} \left(\partial_{\mu}\phi\right)^2\right].
\end{equation}
By applying the scaling transformations, several terms scale zero while the others are invariant (hence marginal), and one obtains the fixed-point action
\begin{equation}\label{fixed-pont1}
    S=\int dt\text{ }d^2x\left[\int^{\infty}_{-\infty} d\chi \bar{\psi}\left(i s \partial_{t} -i v_F \partial_x-i v_F\chi \partial_{y} +g\phi \right)\psi-\dfrac{1}{2} \left(\partial_{y}\phi\right)^2\right].
\end{equation}
where the integral over $\chi$ is now unbounded since if one zooms close to a patch one cannot see the periodicity of the Fermi surface. The fermion kinetic term scales to zero as $s\rightarrow 0$, which reflects the fact that it is subdominant compared to the $\omega^{2/3}$ dynamics generated by the interactions. However, since this highly nonlocal term does not appear in the bare action, it is important to retain the bare kinetic term in order to preserve some form of causality, as well as to ensure that high frequency modes are suppressed. The theory with arbitrarily small kinetic term is fundamentally different from the unphysical theory with no kinetic term. 

It is difficult to apply the above scaling analysis directly to the bosonized action. The bosonized action turns out to contain two naively relevant terms whose effects cancel each other: the bare boson mass and the $\left(\partial_{\perp} \zeta\right)^2$ term. The dominant effect of the latter term is to renormalize the former term; the ``sub-dominant'' effect is to generate  Landau damping. A similar subtlety is that the bare electron-hole propagator appears to violate scaling in momentum space, while in position space it clearly obeys scaling; this is the result of a cancellation of the naively dominant term in the Fourier transform.

The marginal terms in the fixed-point action \eqref{fixed-pont1} seem to depend on two parameters: $g$ and $v_F$. They also implicitly depend on $k_F$ through the cutoff consistency relation, eq. \eqref{uncert}. However, all parameters can be absorbed into the coordinates and the fields, at the cost of changing the coefficient of the irrelevant kinetic term. Thus, the fixed point has no free parameters, and the action can be rewritten as
\begin{equation}\label{fixed-point}
    S=\int dt d^2x\left[\int^{\infty}_{-\infty} d\chi \bar{\psi}\left( i\eta \partial_{t} -i\partial_x-i\chi \partial_{y} +\phi \right)\psi-\dfrac{1}{2} \left(\partial_{y}\phi\right)^2\right].
\end{equation}

In \cite{mehta-talk}, Mehta introduced a different scaling prescription for the bosonized theory, which seems to eliminate the subtle cancellations discussed above. Under Mehta's scaling, $\omega\sim k_x\sim k_y^3\sim k_y \chi$. However, Metlitski and Sachdev argued at low energies, $k_x$ should scale as $k_y^2$ to all orders in the loop expansion. The argument is based on an exact symmetry of the fixed-point action known as sliding symmetry. In the patched theory, this symmetry takes the form
\begin{align}
    \nonumber \chi&\rightarrow \chi+\delta\chi\\
     k_x &\rightarrow k_x-k_y \delta\chi.
\end{align}
This is just a remnant of rotational symmetry in the parabolic Fermi surface, and can be viewed as an infinitesimal rotation with irrelevant terms neglected. In the conventional theory of the unpatched parabolic Fermi surface, which seems to capture the entire physics of the fixed point, this symmetry implies that physical results can depend on $k_x$ only through the invariant $k_x+k_y^2$ ($-k_x+k_y^2$ for the antipodal patch), which can be understood as the distance from the Fermi surface. In this case, $k_x$ will necessarily scale as $k_y^2$.

\section{Breaking and Re-Engineering Bosonization}
\subsection{The Trouble with Bosonization: Quantum Criticality is not Semiclassical}\label{sec:trouble}
In Fermi surface bosonization, one attempts to ``eat the cake and have it,'' defining both position $\boldsymbol{x}$ and momentum $k_F \chi$ simultaneously. According to the uncertainty principle, this necessitates a finite resolution in both $\chi$ and $x_{\parallel}$. Bosonization thus entails a slight nonlocality, reflected by the cutoffs in the case of \textit{patch} bosonization but necessarily present in one form or another in any phase space formalism. However, the nonlocality effectively disappears when the \textit{dynamic} uncertainty in position and momentum is much larger than the kinematic uncertainty, i.e., when the physics is naturally dominated by states that are too delocalized in position and momentum to ``see'' the inherent fuzziness in the definition of $x_{\parallel}$ and $k_F \chi$. The expectation that something like this would happens was the motivation to patch the Fermi surface in the first place. In order for this to happen, the uncertainty principle imposes that  
\begin{equation}
    k_F \delta x_{\parallel\text{ dynamic}} \delta \chi_{\text{dynamic}}\gg k_F \delta x_{\parallel\text{ kinematic}} \delta \chi_{\text{kinematic}}\gtrsim 1.
\end{equation}
This condition describes a separation of length scales between $\delta x_{\parallel\text{ dynamic}}$ and $\frac{1}{k_F \delta\chi_{\text{ dynamic}}}$. This is automatically satisfied in a Fermi liquid at low energies, wherein $\delta \chi_{\text{ dynamic}}\sim 2\pi$ while $\delta x_{\parallel\text{ dynamic}}\sim \omega^{-1}$. In a quantum critical system, however, this doesn't happen. Scale invariance means that any two length scales have to scale in the exactly same way at low energies, and unless some conspiracy is involved, the dimensionless ratio will be some order-one constant. 

In the boson-fermion problem, scaling (Section \ref{sec:scaling}) suggests that  $k_F \delta\chi_{\text{ dynamic}} \sim \omega^{1/3}$ while $\delta x_{\parallel\text{ dynamic}}\sim \omega^{-1/3}$. This can be verified directly from a self-consistent one-loop calculation in the patched theory \eqref{fixed-point}. For the boson self-energy, one finds 
\begin{equation}\label{1l-se}
    \Pi\left(\nu,\boldsymbol{q}\right)\propto \dfrac{1}{d \chi} \int d\chi \int {dk_x dk_y  d\omega} G\left(\omega,k_{\perp},\chi\right) G\left(\omega+\nu,k_{\perp}+q_{\perp},\chi\right),
\end{equation}
with $G$ the full RPA propagator with the $\omega^{2/3}$ self-energy, and the singular factor of $\frac{1}{d\chi}$ comes from the $d\chi$ in the free fermion action and in the interaction term. One can then change integration variables to $k_{\perp}$ and $k_{\parallel}$. Since the linearized fermion propagator is independent of $k_{\parallel}$, the $k_{\parallel}$ integral gives a linear divergence canceling the factor of $d\chi$ by the cutoff consistency relation \eqref{cutoffs}. After $k_{\perp}$-integration, only the imaginary part contributes to the $\chi$-integral. The imaginary part of the $\chi$-integrand, in turn, is a Lorentzian centered around $\chi=q_x/q_y$ with a spread\footnote{In the bosonized calculation, it is slightly more subtle to argue for this, since the contribution to the boson self-energy comes from the infinitely sharp particle-hole pole. To argue for an $\omega^{1/3}$-spread in $\chi$, one must keep in mind that a fermion interacts with an $\omega^{1/3}$ range of bosonic momenta due to Landau damping, and that the contribution to the boson self-energy at different bosonic momenta originates from different patches.} $\delta\chi\sim\frac{\left(\omega+\nu\right)^{2/3}-\omega^{2/3}}{q_y}$, and for typical bosonic momenta (i.e., the ones most strongly contributing to the fermion self-energy) this is about $\nu^{1/3}$. Similarly, for the fermion self-energy
\begin{align}
    \Sigma\left(\omega,\boldsymbol{k},\chi\right)&\propto \int dq_x dq_y d\nu D\left(\nu,q_y\right) G\left(\omega+\nu,k_{\perp}+q_{\perp},\chi\right)\nonumber\\
    &=\dfrac{i \pi}{v_F}  \int dq_y d\nu \text{sign}(\omega+\nu) D\left(\nu,q_y\right).
\end{align}
$D\left(\nu,q_y\right)$ is peaked around $q_y\sim\nu^{1/3}$ with a spread $\delta q_y\sim \nu^{1/3}$. The frequency integral acquires contributions in the range $\left|\omega\right|>\left|\nu\right|>0$, and so the dominant contribution to the fermion self-energy comes from momenta of the order of $\omega^{1/3}$.

In order to correctly calculate the bosonic self-energy, one takes the continuum limit of patches, replacing the sum over patches with an integral. This is only valid when $\delta \chi$ is much wider than the patch spacing, so $\omega^{1/3}\gg \Lambda$. Meanwhile, in order to correctly calculate the fermion self-energy, one must set a cutoff larger than the dominant contribution to the $q_y$-integral, so $\Lambda \gg \omega^{1/3}$. Thus, the patched theory only ``works'' insofar as it treats the cutoff inconsistently!

\subsection{Bosonization as a Small-$N$ Theory}
Patch bosonization is not valid in the boson-fermion model. Nevertheless, in light of the reasonable results and transparent non-Fermi liquid physics it provides, it is natural to wonder whether one can ``cook up'' a limit of the boson-fermion problem in which bosonization is exact, and whether such a limit can teach one anything about more realistic non-Fermi liquids. 

Bosonization assumes that the boson interacts with a range of patches much broader than its typical momentum. The range of patches with which a boson interacts is determined by the fermion self-energy (which broadens the range of patches an electron can occupy), while the typical bosonic momentum is determined by the boson self-energy (which broadens the range of possible momenta a boson can carry). Therefore, a bosonizable limit consists of either singularly suppressed boson damping or singularly enhanced fermion damping. Since the fixed-point action \eqref{fixed-point} has no free parameters, simply adjusting quantities like the Fermi velocity or Fermi momentum will not alter the critical behavior. It is clear that any bosonizable limit will necessarily be artificial and ``violent.'' 

One possibility proposed in \cite{ioffe} is to couple the boson to $N$ fermionic flavors, assume that one can treat $N$ as a continuous parameter and take the formal limit $N\rightarrow 0$. This suppresses Landau damping linearly in $N$ because the boson's decay rate is proportional to the number of decay channels\textemdash however, a quick calculation reveals that the one loop fermion self-energy is \textit{enhanced} as $N^{-1/3}$ due the singularity in the boson's propagator in the absence of damping. Thus $\delta \chi \gg \delta k_y$, and the results of bosonization obtained in Section \ref{sec:bose-res} can be trusted.

Of course, the legitimate goal of any synthetic limit is to teach one something about the real world. This is an especially high bar for the small-$N$ limit, since $N=0$ does not define a model of a non-Fermi liquid\textemdash it is merely an extrapolation (and interpolation) of a parameter in the results of a physical model to an unphysical value. Its usefulness depends entirely on whether the extrapolation is reversible\textemdash i.e., whether the physical $N=1$ problem can be studied via some form of a perturbation theory around the exact $N=0$ solution. This question is not addressed in the original reference.

One can study the crossover to $N=1$ by deriving an exactly solvable bosonized action for $N=0$ and determining how it gets modified as $N$ is varied. The first step is to patch the theory and take a double continuum limit in $\chi$ and in $y$\textemdash a limit which would never be possible in the $N=1$ model. The unpatched action of a parabolic Fermi surface is given by
\begin{align}
    S=\int dt \int d^2x & \left[\sum_{i}\bar{\psi}_{i}\left(i\partial_t-i\partial_{x}-\dfrac{1}{2}\partial_{y}^2+gN^{1/4}\phi\right)\psi_{i}-\dfrac{1}{2} \left(\partial_{y}\phi\right)^2\right]. 
\end{align} 
$gN^{1/4}$ is analogous to the 't Hooft coupling in large-$N$ theories. One might have expected the 't Hooft coupling to be \textit{strong} so as to balance the small number of scattering channels. However, it turns out that the double continuum limit tends to make interactions singularly strong, since it generates a continuum of fermion fields with which the boson can interact; the normalization of the coupling will cancel this effect.

Next, patch the action in the usual way, and rescale $k_y$ and $\chi$ in such a way as to raise the large $k_y$-cutoff and lower the small $\chi$-cutoff:
\begin{align}\label{rescaling}
    k_y&\rightarrow \sqrt{N} k_y \nonumber\\
    \chi&\rightarrow \dfrac{\chi}{\sqrt{N}},
\end{align}
so that in terms of the original cutoff $\Lambda_0$, the new cutoffs become
\begin{align}\label{rescaled-cutoffs}
    \Lambda_{k_y}&=\dfrac{\Lambda_0}{\sqrt{N}} \nonumber\\
    \Lambda_{\chi}&=\Lambda_0 \sqrt{N}.
\end{align}
This will not change the fermion dispersion $\partial_x+\chi\partial_y$; however, it will transform the measure $dy d\chi$ of the fermion action, and the free fermion term will change as
\begin{equation}
    S_f\rightarrow \dfrac{S_f}{N}.
\end{equation}
Since the fermion action is quadratic in $\psi$, this change can be absorbed into $\psi$ via
\begin{equation}\label{rescaled-psi}
    \psi\rightarrow \sqrt{N}{\psi}.
\end{equation}
Furthermore, due to $k_y$-scaling, the boson action acquires a prefactor
\begin{equation}
    S_{b}\rightarrow \sqrt{N}{S_b},
\end{equation}
which can be reabsorbed into the boson field
\begin{equation}
    \phi\rightarrow \dfrac{\phi}{N^{1/4}}.
\end{equation}
The effect is to change the interaction term
\begin{equation}
    S_{int}\rightarrow \dfrac{1}{N^{1/4}} S_{int},
\end{equation}
so in the rescaled theory, the coupling is independent of $N$, and the patched action takes the form 
\begin{equation}
    S=\int_{\frac{\Lambda_0}{\sqrt{N}}} d^2 r\text{ }dt\left[\int_{{\Lambda_0}\sqrt{N}} d\chi \sum^{N}_{i=1} \bar{\psi}_{i}\left(i\partial_t-\partial_{x}-\chi \partial_y+g \phi\right)\psi_{i}+\dfrac{1}{2} \phi \partial_y^{2} \phi \right], 
\end{equation}
with the subscripts of the integral symbols representing cutoffs.

Of course, this rescaling is just a redefinition of variables, which can always be done and does not change the physics. The promise of the small-$N$ limit is to render physical results non-singular even in the limit of singular scaling. At the one-loop level, this promise is fulfilled: the boson's self-energy will get a factor of $N$ due to the $N$-fermions to which it can decay\textemdash but recall from \eqref{1l-se} that the boson's self-energy also contains a divergent-looking factor of $\frac{\Lambda_{k_y}}{d\chi}$; in the original model, this factor is just unity, but due to the rescaled cutoffs, it will instead be given by $\frac{1}{N}$. The one-loop fermion self-energy has no additional factors of $N$, so the one loop Green's functions are independent of $N$. 

Beyond one loop, the rescaled small-$N$ theory can be exactly solved by bosonization. However, one must keep in mind that the prefactor of the $\zeta$-action is always one when $y$ and $\chi$ are discretized with the appropriate cutoffs. Thus, using $d\chi dy=\frac{1}{N}$, the bosonized action will take the form
   \begin{align}\label{small-n-bose}
    S[\zeta,\phi]=&\dfrac{k_F}{8\pi^2} N \sum^{N}_{i=1} \int_{\sqrt{N}\Lambda_0} d\chi  \int_{\frac{\Lambda_0}{\sqrt{N}]}} dt\text{ } d^2 r \partial_{\perp} \zeta_{i}\left\{\left(\partial_{t}-\partial_{\perp} \right) \zeta_{i}+2g \phi \right\}\nonumber\\
    & -\dfrac{1}{2} \int_{\frac{\Lambda_0}{\sqrt{N}}} dt\text{ }d^2 r\left(\partial_y \phi\right)^2.
\end{align}
The prefactor of $N$ can be absorbed into $\zeta$, which will once again rescale the coupling. This rescaling is consistent with the fermionic theory. The rescaling of $\zeta$ will change its periodicity, which must be kept in mind when calculating fermion propagators.

As one varies $N$ from 0 to 1, the cutoffs on $\chi$ and $k_y$ will shift from zero and infinity to the same finite number $\Lambda_0$. This will have two effects:
\begin{enumerate}
    \item The integrals over $\chi$ and $y$ will do an increasingly poor job of approximating the appropriate discrete sums.
    \item As the flat patches become wider and wider, they will do an increasingly poor job of approximating the shape of the Fermi surface.
\end{enumerate}
To account for the intrinsic curvature of a patch, one can add a quadratic term to the fermion dispersion in the unscaled action, turning the patches into appropriately curved parabolas:
\begin{equation}
    S=\int_{\Lambda_0} d^2 r\text{ }dt\left[\int_{\Lambda_0} d\chi \sum_{i} \bar{\psi}_i\left(i\partial_t-\partial_{x}-\chi \partial_y-\dfrac{1}{2}\partial_{y}^2 +\dfrac{g}{N^{1/4}} \phi\right)\psi_{i}+\dfrac{1}{2} \phi \partial_y^{2} \phi \right]. 
\end{equation}
I will call the new term the intra-patch curvature to distinguish it from the inter-patch curvature $\chi \partial_{y}$ which reflects the fact that different patches have different normals. By power counting, this term is marginal, while all higher-order terms in the dispersion (including $\partial_x^2$) are irrelevant. Under rescaling, the action takes the form
\begin{equation}
    S=\int_{\frac{\Lambda_0}{\sqrt{N}}} d^2 r\text{ }dt\left[\int_{\sqrt{N}\Lambda_0} d\chi \sum_{i} \bar{\psi}_i\left(i\partial_t-\partial_{x}-\chi \partial_y-\dfrac{N}{2}\partial_{y}^2 +{g} \phi\right)\psi_{i}+\dfrac{1}{2} \phi \partial_y^{2} \phi \right]. 
\end{equation}
so at small $N$, one should be able to treat the intra-patch curvature as a perturbation. When bosonized, it will be shown that it turns into a cubic $\zeta$-vertex (with some caveats).

By contrast, there can be no perturbation theory describing the crossover from integrals to sums and from the multi-patch theory to the single-patch theory. The perturbative expansion of a discrete sum near its continuum limit is given by the Euler-Macluarin formula:
\begin{equation}
    \sum^{m+1}_{i=n} f(i\delta x)=\int^{m\delta x}_{n\delta x} f(x) dx+\sum^{N}_{k=1} \left(\delta x\right)^{k} \dfrac{B_k}{k!}\left(f^{(k-1)}(n\delta x)-f^{(k-1)}(m\delta x)\right)+\mathcal{O}\left(\delta x^{N+1}\right),
\end{equation}
with $B_k$ the Bernoulli numbers. The corrections are differences between derivatives at the two boundaries of the integral. In the case of interest, all the integrals are improper, and all the boundaries are sent to infinity, in which case the boundary terms vanish individually. Alternatively, if one wishes to keep track of the ``UV'' periodic behavior of $\chi$, or to confine the system to a finite volume with periodic boundary conditions in $y$, then the two boundaries are identified and all the correction terms cancel. Yet the integral is clearly not equal to the sum! The conclusion is that the difference between the sum and the integral must be smaller than any power of $\delta x$ as $\delta x\rightarrow 0$, as exemplified by the function $e^{-\frac{1}{\delta x^2}}$. This conclusion is confirmed by playing with discretized integrals on Matlab.

Consequently, the continuum theory with intra-patch curvature is practically \textit{exact} in a range of sufficiently small but non-vanishing values of $N$. However, when $N$ is sufficiently large, non-perturbative effects become important. Hence, there is an unexpected crossover to discrete-patch behavior that cannot be extrapolated from the continuum limit. The small-$N$ perturbation theory might still converge at this point, but it must converge to the \textit{wrong answer}. This argument applies beyond the small-$N$ limit: it assumes only that one starts from a continuum of patches with intrinsic curvature and gradually increases their width; the small-$N$ limit is just one way to make this process non-singular. It is tempting to conclude that this applies \textit{any} bosonizable limit\textemdash i.e., that the linear bosonized theory is not perturbatively connected to the single patch fixed point. The perturbation theory proposed in \cite{mehta} based on nonlinear bosonization might seem like a loophole, since in nonlinear bosonization there are no sharp cutoffs on momentum; however, I will argue in the discussion that this perturbation theory is singular.

\subsection{Bosonization as a Large-$N$ Theory}
The small $N$ theory is far from fully satisfying because $N=0$ is a trivial state with no fermions, and the formal extrapolation to this state is not a genuine example for a bosonizable non-Fermi liquid.  I will propose a \textit{large}-N model that is exactly bosonizable for a similar reason. It is the \textit{only} bosonizable model for a two-dimensional non-Fermi liquid I have been able to find\footnote{This includes the Nayak-Wilczek limit that was considered in the context of coadjoint orbit bosonization in \cite{mehta} with the claim that it makes nonlinearities weakly relevant; I found the inconsistency of linear bosonization remains intact in this limit.}. However, unlike the small-$N$ theory, the large-$N$ theory produces a considerably more complicated bosonized action than eq. \eqref{bosonized-action}.

Assume a large number of fermion flavors $n$ coupled to a much \textit{larger} number of bosonic flavors $N\gg n$. The rate of decay of the fermions (bosons) will be controlled by the size of the bosonic (fermionic) bath; the idea is to choose the ratio between $N$ and $n$ to be large enough that the fermion Green's function becomes much broader than the boson Green's function. Note that one cannot simply take $n=1$, since when $n=1$ the fermions couple to a single linear combination of bosons, which is equivalent to the original problem up to a large number of decoupled modes. Preventing this also requires that the coupling constant be flavor-dependent. In order for the interaction to be bosonizable, the boson must couple to fermion charge densities (with no inter-flavor tunneling terms of the form $\overline{\psi}_{i}\psi_{j}$). The most general interaction term is thus
\begin{equation}
    S_{int} = \sum^{n}_{i=1}\sum^{N}_{J=1} \int d^2 x\text {}dt g_{iJ} \left|\psi_{i} \left(\boldsymbol{x},t\right)\right|^2 \phi_{J}\left(\boldsymbol{x},t\right)
\end{equation}
with uppercase letters representing bosonic indices and lowercase letters representing fermionic indices. In general, this action will allow an individual boson to change flavor upon emitting and reabsorbing a fermion, a scenario not possible in the bare action\textemdash this will lead to a $t$-matrix-like term in the boson propagator (similarly to the one found for the electron-hole pair in the bosonized theory), which will complicate matters at best and significantly change the physics at worst. I will escape this worry using Sachdev's trick of treating the couplings as  uncorrelated, Gaussian random variables, with
\begin{align}
    \overline{g_{iI}}&=0\nonumber\\
    \overline{g_{iI}g_{jJ}}&=g^2 \delta_{ij}\delta_{IJ}.
\end{align}
The fact that the $g_{iI}$s are uncorrelated ensures that the amplitude for any process will vanish unless every flavor exchange is reversed, and the boson's flavor is conserved on average. This model differs from the random large-$N$ theory of Esterlis et al.\cite{Sachdev} in that the interaction does not mix different fermionic flavors.

Bosonization is achieved much like in the small-$N$ case. One finds that the one-loop boson self-energy is proportional to $n$, while the one-loop fermion self-energy is proportional to $\frac{N}{n^{1/3}}$. This dependence on $n$ and $N$ can be eliminated by introducing the 't Hooft coupling
\begin{equation}
    g_{iJ}\rightarrow g_{iJ}\dfrac{n^{1/4}}{N^{3/4}},
\end{equation}
and then rescaling
\begin{align}
    k_y&\rightarrow \sqrt{\dfrac{n}{N}} k_y\nonumber\\
    \chi&\rightarrow \sqrt{\dfrac{N}{n}} \chi.
\end{align}
In the small $\frac{n}{N}$ limit, this rescaling is the double continuum limit. The action, rescaled as in the small-$N$ theory, takes the form
\begin{align}
    S=\int_{\Lambda_0\sqrt{\frac{n}{N}}} d^2 r\text{ }dt\left[\int_{\Lambda_0\sqrt{\frac{N}{n}}} d\chi \sum^{n}_{i=1} \bar{\psi}_{i}\left(i\partial_t-\partial_{x}-\chi \partial_y+\sum^{N}_{J=1} \dfrac{g_{iJ}}{\sqrt{N}} \phi_{J}\right)\psi_{i} \right.& \nonumber\\
    \left. +\dfrac{1}{2} \sum^{N}_{I=1}\phi_{I} \partial_y^2 \phi_{I}\right] &.
\end{align}
Bosonizing this and rescaling $\zeta$ as in the small-$N$ theory gives
\begin{align}\label{large-N-bose}
    S[\zeta,\phi]=&\dfrac{k_F}{8\pi^2}  \sum^{n}_{i=1} \int_{\sqrt{\frac{n}{N}}\Lambda_0} d\chi  \int_{\sqrt{\frac{N}{n}}\Lambda_0} dt\text{ } d^2 r \partial_{\perp} \zeta_{i}\left\{\left(\partial_{t}-\partial_{\perp} \right) \zeta_{i}+\sum^{N}_{J=1}\frac{2 g_{iJ}}{\sqrt{n}} \phi_J \right\}\nonumber\\
    & +\sum^{N}_{I=1}\dfrac{1}{2} \int_{\sqrt{\frac{N}{n}}\Lambda_0} dt\text{ }d^2 r\left(\partial_y \phi_I\right)^2.
\end{align}

For a specific realization of $g_{iJ}$, the bosonized action is quadratic and thus the moment generating function can be formally computed. However, the result depends on the inverse of a random $n+N$ by $n+N$ matrix. Averaging over $g_{iJ}$ results in a quartic action, which cannot be solved exactly. Nevertheless, the theory remains, in a sense, exactly solvable: in particular, the boson's self-energy series, averaged over $g_{iJ}$, remains a series of single-body diagrams that can each be evaluated exactly. It is easy to see that every term has the same energy- and momentum dependence as in the original bosonized theory, and that only the prefactors get modified, so one finds\footnote{There is a subtlety I am sweeping under the carpet here. In the bosonized theory $\phi$ always acquires a mass renormalization. In the random coupling theory, the mass renormalization becomes a matrix proportional to $\sum_{k} g_{kI} g_{kJ}$, so canceling it at each order of the self-energy series requires a random bare (squared-)mass matrix distributed as (the square of) a normal variable.}:

\begin{equation}\label{rand-prop}
    D(q,\omega)=\sum^{\infty}_{n=0} A_{n}\dfrac{1}{q_{y}^{2+2n}} \dfrac{\left|\omega\right|^{n}}{\left|q_y\right|^{n}}=\dfrac{1}{q^2_{y}+ F\left(\dfrac{\left|\omega\right|}{\left|q_y\right|^3}\right)\dfrac{\left|\omega\right|}{\left|q_y\right|}},
\end{equation}
for some non-singular function $F$ that can be expressed as a series. While the functional form of the boson propagator changes, the critical exponents do not. A similar conclusion applies to the fermion propagator.

\subsection{Structure of the Expansion Around the Bosonized Solution}
Like the small-$N$ expansion, the double-large $N$ expansion is a perturbation theory in the intra-patch curvature. As before, the expansion can never converge to the correct answer at $n=N=1$. Nevertheless, it might still describe a class of controllable ``toy models'' for non-Fermi liquids, and the extrapolation of its results to $n=N=1$ might still be an uncontrolled approximation to the realistic non-Fermi liquid, so it is of some interest to understand it. I have not calculated any $n/N$-corrections since the calculation involves a large number of complicated integrals even at the leading nontrivial order. Rather than describing any detailed calculation, I will just outline the general structure of the proposed expansion, to the extent that I understand it, and some subtleties that must be taken into account in order to complete the program.

The intra-patch curvature term is given by
\begin{equation}
    S_{\gamma}={\gamma} \int d^2 x\text{ }dt \int^{2\pi}_0 d \chi \partial_{y} \bar{\psi}(\chi) \partial_{y} \psi (\chi).
\end{equation}
where $\gamma=\frac{n}{2N}$ in the double-large $N$ theory. I will omit flavor indices and cutoffs for notational brevity. Bosonizing this term will lead to a cubic vertex that one might hope to treat as a perturbation. This is most easily achieved in the Hamiltonian picture, where one just needs to plug in the relation between $\psi$ and $\zeta$, $\psi\sim e^{i\zeta}$.

However, this naive relation between $\psi$ and $\zeta$ ignores the fact that, while the $\zeta$ fields commute at different values of $\chi$ and $x_{\parallel}$, the $\psi$s must certainly anti-commute. In the linearized theory, one didn't have to worry about this because the \textit{dynamics} never exchanged two fermion fields: since the fermions of each patch could only move in the $x_{\perp}$-direction, like cars in a single-lane road, they were unable pass each other. (The boson-fermion vertex cannot permute two fermions either, as it always creates a \textit{localized} electron-hole pair.) This is precisely the sense in which fermions and bosons are equivalent in one-dimensional systems. By contrast, the intra-patch curvature allows fermions hop in the $x_{\parallel}$-direction. Thus, two fermions can now easily hop around each other and thereby exchange their positions. Amplitudes must gain a phase of $-1$ whenever this happens, and there must be an extra phase factor accounting for this. Note that relative phases of $-1$ between \textit{different} states in Fock space are a gauge ambiguity of fermions. The only physical statement is that any \textit{closed loop} in Fock space gives a phase of $-1$ if it involves an odd number of permutations of fermions; \textit{any} choice of a phase factor which achieves this is equivalent, much like the equivalence of different choices of the vector potential that produce the same magnetic field.

One possible choice of the phase factor can be obtained by thinking of the Fermi surface as an infinite collection of one-dimensional wires labeled by $x_{\parallel}$ and $\chi$, and introducing fermion operators that keep track of the fermion parity of each wire:
\begin{equation}\label{rel}
    \psi^{\dagger}(t,\boldsymbol{r},\chi)=\dfrac{1}{\sqrt{2\pi}} \eta\left(t,x_{\parallel},\chi\right) \normord{\exp\left[i\zeta\left(t,\boldsymbol{r},\chi\right)\right]}.
\end{equation}
where the double colon represents normal ordering. $\eta$ are nonlocal Fermi fields which do not depend on $x_{\perp}$. They should be taken to be Majorana fermions to ensure that creating two fermions in the same wire does not annihilate the state, so that $\eta=\eta^{\dagger}$ and $\eta^2=1$. These fields are known as \textit{Klein factors}; see \cite{sech} and \cite{begin} for an introduction in the one-dimensional case.

Plugging the fundamental relation \eqref{rel}, and using point-splitting to regulate the divergence of products of operators at the same point, one finds
\begin{align}\label{expr}
    \partial_{y} \psi^{\dagger}(\boldsymbol{r}) \partial_{y} \psi (\boldsymbol{r})=& \dfrac{1}{2\pi} \lim_{\boldsymbol{r}\rightarrow{\boldsymbol{r^\prime}}}{{\partial_{y}\left\{\eta\left(y\right)\normord{\exp\left[i\zeta(\boldsymbol{r})\right]}\right\}\partial^{\prime}_{y}\left\{\eta(y^\prime) \normord{\exp\left[-i\zeta\left(\boldsymbol{r^\prime}\right)\right]}\right\}}} \nonumber\\
    =&\dfrac{1}{2\pi}\lim_{\boldsymbol{r}\rightarrow{\boldsymbol{r^\prime}}}{{\normord{\partial_{y} \zeta \left(\boldsymbol{r}\right) \exp\left[i\zeta(\boldsymbol{r})\right]} \normord{ \partial^{\prime}_{y} \zeta \left(\boldsymbol{r^\prime}\right)\exp\left[-i\zeta\left(\boldsymbol{r^\prime}\right)\right)]}}}\nonumber\\
    &+\dfrac{1}{2\pi}\partial_{y}\eta \lim_{\boldsymbol{r}\rightarrow{\boldsymbol{r^\prime}}}{{\normord{\partial_{y} \zeta \left(\boldsymbol{r}\right) \exp\left[i\zeta(\boldsymbol{r})\right]} \normord{  \exp\left[-i\zeta\left(\boldsymbol{r^\prime}\right)\right)]}}}\nonumber\\
    &-\dfrac{1}{2\pi}\partial_{y}\eta \lim_{\boldsymbol{r}\rightarrow{\boldsymbol{r^\prime}}}  {{\normord{ \exp\left[i\zeta(\boldsymbol{r})\right]} \normord{\partial^\prime_{y} \zeta\left(\boldsymbol{r^\prime}\right)  \exp\left[-i\zeta\left(\boldsymbol{r^\prime}\right)\right)]}}}.  
\end{align}
There is, in principle, also a $\left(\partial_{y} \eta\right)^2$-term that has been dropped, but it is just a shift of the chemical potential by the Majorana property. One can then decompose $\zeta$ into a sum of creation operators and annihilation operators, $\zeta=\zeta_{+}+\zeta_{-}$, and bring all the creation operators to the left of the annihilation operators using the algebra of the fields:
\begin{equation}
\left[\zeta_+\left(\boldsymbol{r},\chi\right),\zeta_-\left(\boldsymbol{r^\prime},\chi^\prime\right)\right]=2\pi^2\delta\left(\chi-\chi^\prime\right)\delta\left(x_{\parallel}-{x^\prime_\parallel}\right) \text{sign}\left(x_{\perp}-x^\prime_{\perp}\right),
\end{equation}
and using the fundamental identity of bosonization: 
$\normord{\psi^{\dagger} \psi}=\frac{1}{4\pi^2}\partial_{\perp}\zeta $.
The result is 
\begin{equation}\label{bose-intra}
    H_\gamma=\dfrac{\gamma}{4\pi^2} \int^{2\pi}_0 d\chi \int d^2 x \left[  \normord{\left(\partial_{y} \zeta\right)^2 \partial_{\perp} \zeta} +2\eta \partial_{y} \eta\normord{ \partial_{\perp} \zeta  \partial_{y} \zeta}\right].
\end{equation}
One can translate this into a path integral for $\zeta$ and for the Klein fields straightfowardly (this is the advantage of the \textit{chiral} version of bosonization). The Klein fields do not seem to affect the controllability  of the expansion, but they do destroy the purely bosonic interpretation of the Fermi surface if they are relevant at low energies.

The small $\gamma$ expansion is thus an expansion in the cubic and quartic vertices in \eqref{bose-intra}, but with the \textit{fully dressed} Gaussian propagators that were computed throughout Section \ref{sec:bose-res}. There are reasonable grounds to suspect that this expansion is plagued with UV divergences. Recall that the fully dressed Gaussian electron-hole propagator \eqref{eh-prop} contains a patch-local term. Due to the patch locality of the \textit{vertex}, this forces a class of loop diagrams to be fully patch-local. In such diagrams, one can change the loop integration variables from $k_x$ and $k_y$ to $k_{\perp}=k_x+\chi k_y$ and $k_{\parallel}=k_y$; the free electron-hole propagators then do not depend on $k_{\parallel}$, but the vertices depend on $k_{\parallel}$ polynomially. Thus, some of these diagrams will diverge as a power-law in the UV (depending on the number of $\phi$-propagators). Such divergences will lead to a breakdown of scaling, and considering the dependence of the cutoffs on $n/N$, it will change the dependence of diagrams on the small parameter\textemdash unless the divergences cancel.

These possible divergences are intuitive. Analogous divergences arise in the attempt to treat the kinetic term in the Schrödinger equation as perturbation to the potential energy. The $\gamma=0$ theory has no term that energetically penalizes  configurations of $\zeta$ which are discontinuous in $x_{\parallel}$. Thus, the linearized theory is dominated by them; but the new term gives those term an infinite energetic cost, so the original assumption that it is ``negligible'' is inconsistent. However, there is exactly one additional quadratic, local term that is consistent with all the symmetries of the action (e.g., parity, time reversal, patch-dependent phase shift) and that does not change the scaling of the Gaussian theory. This term is
\begin{equation}\label{c-term}
    S_{C}= C\int dt\text{ }d^2 x\int^{2\pi}_{0}\text{ }d\chi  \left(\partial^2_{y} \zeta \right)^2.
\end{equation}
This term energetically suppresses high-$k_{\parallel}$-modes. Since the theory remains exactly solvable with it, the $\gamma$-perturbation theory should be tractable and non-singular once this term is included. The idea is that since there is no obvious reason for the $C$-term to be forbidden,  the $\gamma$-vertex may dynamically generate it. If this happens, then one can start from a quadratic theory with some unknown, bare $C_0(\gamma)$, and then at each order in $\gamma$ tune $C(\gamma)$ so that the renormalized value of $C$ is the same as it would be in the physical case $\gamma=\frac{1}{2}$ (with no bare $C_0$). This version of perturbation theory should be controllable. In fact, if the diagrams that renormalize $C$ themselves diverge in the naive perturbation theory, then the renormalized value of $C$ is non-zero even in the simple-minded $\gamma \rightarrow 0$ limit, and it is never consistent to drop it\footnote{An additional attractive feature of the $C$-term is that it renders the interaction with the Klein field irrelevant. The Klein fields describe the fluctuations the total charge of the wire labeled by $x_{\parallel},\chi$. In fact, the Klein factors can be replaced by a Wigner-Jordan phase depending on the zero mode of the charge density; see \cite{sech}. But since the total charge of a wire is proportional to the winding number of the phase $\frac{1}{2\pi}\int^{\infty}_0 dx_{\perp} \partial_{\perp} \zeta $, which is quantized when one imposes periodic boundary conditions, fluctuations of this quantity require field configurations discontinuous in $x_{\parallel}$, which are infinitely penalized by the $C$-term.}.

\begin{figure}
    \centering
    \includegraphics[width=0.5\linewidth]{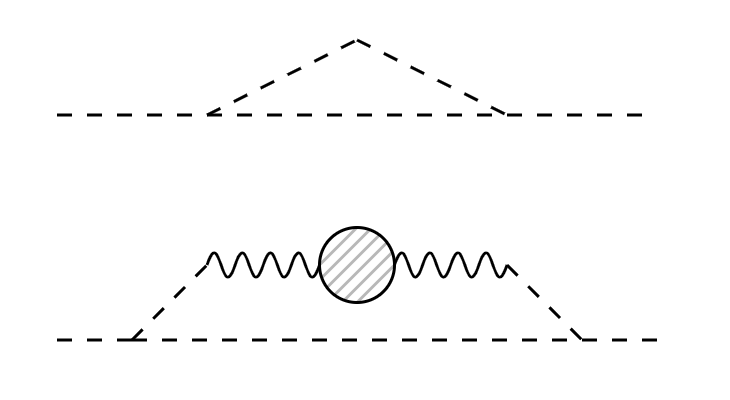}
    \caption{The two patch-local (i.e., initial $\chi$ equal to final $\chi$) corrections to the electron-hole propagator to order $\gamma^2$. Dashed lines are bare electron-hole propagators. The wavy line with the filled bubble is a full \textit{linearized} $\phi$-propagator from \eqref{boson-prop}, with all corrections from $g$ but no corrections from $\gamma$.  The upper diagram vanishes. The lower diagram does not generate the $C$-term from eq. \eqref{c-term}.}
    \label{fig:gamma}
\end{figure}

I considered the diagrams contributing to the electron-hole propagator at naive order $\gamma^2$, depicted in figure \ref{fig:gamma}, while ignoring the Klein fields and the modification of the propagators due to the random couplings (eq. \eqref{rand-prop}). I found that neither of the two patch-local diagrams that arise at this order generate the $C$-term, though one of them turns out to vanish, and I did not calculate the other. It is clear that more work is required to determine whether the perturbation theory around linearized bosonization is sensible.

\section{Discussion and Implications for Other Work}\label{sec:discussion}
\subsection{Coadjoint Orbit Bosonization}
It has been argued that the cutoffs intrinsic to patch bosonization are too tight to accommodate the range of momenta relevant for the critical metal. An alternative to patch bosonization which has been claimed to eliminate the need for cutoffs is coadjoint orbit bosonization (henceforth COB) \cite{dv}\cite{dv2}\cite{sohn}\cite{mehta}\cite{dv3}. COB has been used to successfully reproduce results of Fermi liquid theory beyond linear response. COB may thus achieve a wider range of applicability than linearized patch bosonization by accounting for more irrelevant corrections to the Fermi liquid fixed point. However, at the lowest energies of a Fermi liquid, the two approaches appear to be equivalent. As will be seen, the equivalence is not fully established due to some fundamental open questions about COB. 

COB is based on the conjecture that the space of low energy states of electrons in a metal is spanned by local deformations of the Fermi surface. Any such deformation can be represented as a canonical transformation acting on the equilibrium Fermi sea, parameterized by a single function of phase space $\Phi(\boldsymbol{x},\boldsymbol{p})$ (e.g., imagine evolving the system under a Hamiltonian $\Phi(\boldsymbol{x},\boldsymbol{p})$ over unit time). Since this representation of the Fermi surface is redundant (i.e., two different canonical transformations can result in the same Fermi surface), one can reduce the dependence of $\Phi$ on momentum $\boldsymbol{p}$ to a single angle $\chi$.

COB proposes an effective action for $\Phi$ that gives the collisionless Botlzmann equation as an equation of motion. The action necessarily contains an infinite hierarchy of nonlinear, higher derivative terms, represented as a power series in the Poisson bracket $\frac{1}{k_F}\left( \overset{\leftarrow}{\partial_\chi} \overset{\rightarrow}{\partial_{\parallel}}-\overset{\leftarrow}{\partial_\parallel} \overset{\rightarrow}{\partial_{\chi}}\right)$. Thus, what replaces the sharp cutoffs is fuzziness over small values of $x_{\parallel}$ and $\chi$ in accordance with the uncertainty principle. In a Fermi liquid, higher derivatives can be discarded as irrelevant. One is left with a local action which turns out to be precisely the same as the one obtained in patch bosonization, though seemingly without cutoffs. In the critical metal, $\partial_\chi$ scales to infinity as fast as $\partial_\parallel$ scales to zero (see Section \ref{sec:trouble}), so all terms are equally important at low energies. It is to be concluded that the bosonized action of the critical metal is highly nonlocal and non-polynomial (in violation of the original point of the semiclassical phase-space variables).  Nevertheless, Mehta proposed an expansion for non-Fermi liquids which treats the nonlinearities as a perturbations around cutoff-free, linearized theory\cite{mehta}\cite{mehta-talk}.

However, the absence of cutoffs in COB raises questions about the foundations of the formalism. In the linearized theory, each state is specified by a \textit{continuum of field  configurations} $\Phi(\boldsymbol{r},\chi)$, depending on $d+1$ coordinates, even though one has started from a microscopic theory that has a \textit{single} $d$-dimensional field $\psi(\boldsymbol{r})$. The phase space variables of COB are supposed to emerge from the Wigner transform of the fermion bilinear $T(\boldsymbol{x},\boldsymbol{y})=\psi^{\dagger}(\boldsymbol{x})\psi(\boldsymbol{y})$. Yet the fermion bilinear is a highly constrained quantity: by definition, it must satisfy an infinite set of constraints $T(\boldsymbol{x},\boldsymbol{y})T(\boldsymbol{z},\boldsymbol{s})=-T(\boldsymbol{x},\boldsymbol{s})T(\boldsymbol{z},\boldsymbol{y})$ up to contact terms; no analogue of these constraints is explicitly imposed in COB. It thus seems like Hilbert space is infinitely expanded. This would have measurable consequences\textemdash for example, the specific heat is proportional to the number of degrees of freedom that can absorb energy, and will diverge if there is a true continuum of independent fields (a problem recognized by Delacaratez et al.\cite{sohn} and by Mehta\cite{mehta}, who state that they found similar divergences in other calculations). 

Where have all these extra degrees of freedom come from?\textemdash In patch bosonization, the answer is obvious: The additional fields are simply the high-momentum modes that  have been cut off from $k_{\parallel}>\Lambda$. The cutoff consistency relation \eqref{uncert} ensures that the number of degrees of freedom remains unchanged, and that the specific heat is finite\footnote{In \cite{mehta}, Mehta claims that it is sufficient to impose a UV cutoff on momenta of the order of $k_F$, without a discretization of $\chi$, to regulate the divergence and obtain the correct coefficient of the specific heat of a Fermi liquid. I believe that this claim is founded on a technical error. To obtain the specific heat Mehta calculates the partition function, which in a Gaussian theory is the product of eigenvalues of the inverse propagator. Hence, the free energy will be a discrete \textit{sum} of the logarithms of eigenvalues\textemdash i.e., a sum over $k_x$, $k_y$, $\chi$ and Matsubara frequencies. The sum over momenta $\boldsymbol{k}$ and frequencies can be traded for an integral with an infinitesimal measure $dk_x dk_y d\omega$, at the cost inserting a factor of the system volume $V$ and the temperature $T$. However, the sum over angles $\chi$ remains measureless, and will diverge in the continuum limit of patches. This last fact was missed in the calculation.} \footnote{In the small $N$ and double large-$N$ theories, the rescaling of $x_{\parallel}$ and $\chi$ will reduce the volume of the system and the range of values of $\chi$ as measured in the rescaled units. This reduction will cancel the dependence of the specific heat on the cutoff and eliminate the divergence. Of course, in order to obtain this cancellation one must keep track of the cutoffs throughout the calculation.}. In COB, the question is more difficult to answer, since there is apparently no known microscopic derivation of the COB effective action\footnote{Park et al. attempted to microscopically derive the action of COB\cite{park}. However, the ``effective action'' they derived is the ``quantum effective action'' (Legendre transform of the generating functional), which one extremizes to compute the expectation value of the fields; this action is completely distinct from the ``Wilsonian effective action'' that appears in the path integral, as the one postulated in COB.}. However, one relevant clue is that the kinetic term of COB is linear in time derivatives, and has the schematic form
\begin{equation}
   S_{\text{kinetic}}=\int d^2 r dt d\chi F\left[\Phi;\boldsymbol{r},t,\chi\right]\partial_{t} \Phi
\end{equation}
with $F$ a nonlinear functional of $\Phi$, which is also a function of position, containing spatial and angular derivatives of arbitrary orders but no time derivatives. By differentiating the action with respect to $\partial_{t}\Phi$, one finds that the canonical momentum of $\Phi(\boldsymbol{r},t,\chi)$ is $F\left[\Phi;\boldsymbol{r},t,\chi\right]$. Since this depends in a complicated way on $\Phi$ at all other points and angles, it is to be concluded that the $\Phi$s are not independent degrees of freedom and that a pair of $\Phi$s does not commute.

Thus, the nonlinearities of COB have a crucial kinematical effect in addition to their dynamical one: they specify the Hilbert space of the system. \textit{If} COB is consistent, then this kinematical effect should be enough to constrain the dimension of Hilbert space to be no greater than that of $d$-dimensional functions $\psi(\boldsymbol{r})$, and to render the specific heat finite. Clearly, this would mean that the effect of the nonlinear terms is non-perturbative, since there is no perturbative expansion of a finite specific heat around an infinite specific heat.

Let me conclude the discussion of COB with the general remark that relying on a purely effective field theory reasoning seems dangerous when the degrees of freedom of the effective theory are defined through a highly nonlocal transformation of the microscopic degrees of freedom, and in situations where the derivative expansion clearly breaks down. For example, in patch bosonization it would be impossible to guess that the highly nonlocal Klein factor terms obtained from the intra-patch curvature are ``allowed,'' if not for the microscopic derivation showing that they are \textit{necessary}. Microscopic derivations will likely be an important step toward a complete understanding of the COB formalism.

\subsection{Exact Results From Anomalies}
The bosonic gap obtained by bosonization in the $k\rightarrow 0$ limit of \eqref{boson-se} completely agrees with the exact results obtained directly from the chiral anomaly of the patch $U(1)$ symmetry by Shi et al.\cite{goldman}. The authors consider a model with a patched Fermi surface and intra-patch curvature, 
\begin{align}
    S=\int^{2\pi}_0 d\chi \int dt\text{ }d^2x \bar{\psi}\left(i\partial_t-i v_F\partial_{\perp}-\dfrac{1}{2m}\partial_{\parallel}^2+g\phi\right)\psi+S[\phi], 
\end{align} 
but their results do not depend on the choice of cutoffs and on the $\partial_\parallel^2$ term. These results were used to argue that the optical conductivity of the simple version of the boson-fermion model is the same as that of free fermions. 

The consistency with bosonization is not surprising: while bosonization cannot be applied to the critical, non-Fermi liquid regime, defined by $\omega\ll v_F k$, it is exact in the \textit{transport} regime, $\omega \gg v_F k$, in which each mode of the boson interacts with the entire Fermi surface rather than with a single patch, and so $\delta \chi\sim 2\pi$ while $\delta k_{\parallel}$ is small. In the simple boson-fermion model considered throughout this paper, bosonization predicts that the transport regime is unaffected by the critical regime, since in a Gaussian theory different momenta are decoupled. In models with multiple flavors that can mix by interacting with the boson (i.e., models with a coupling of the form $\phi \bar{\psi}_{i} \psi_{j}$), the bosonized action will not be Gaussian, and the conclusion for the optical conductivity does not follow. This is consistent with a subsequent work by the same authors \cite{goldman2}, where it was found that in certain random coupling large-$N$ models the optical conductivity has a critical contribution. 

\subsection{Bosonization as an Uncontrolled Approximation}
One can get rid of the $k_y$-cutoff of bosonization by  keeping patches explicitly discrete and making them infinitely wide, so that in the end one is left with a single patch (and its antipode), which should be enough to account for the critical regime of the boson-fermion problem. Clearly, the intra-patch curvature becomes singularly relevant in this case, since all curvature is intra-patch when there is one patch! In the absence of the intra-patch curvature, the theory is not well-behaved, and so there is no hope of treating the intra-patch curvature as a perturbation; since bosonization turns the intra-patch curvature from a quadratic term into a cubic vertex, the bosonized theory becomes even less controllable than the fermionic theory. 

Nevertheless, it is reasonable to \textit{assume} that the cubic vertex gives rise to a Landau-damped bosonic propagator, as obtained in multi-patch linearized bosonization and in the Eliashberg theory. Then one can sum over all diagrams of the \textit{Gaussian} part of the action, but with the \textit{full} conjectured boson propagator, to obtain an approximate expression for the fermion propagator. This expression should be identical to the one obtained in multi-patch bosonization, \eqref{ferm-prop}. While this approximation is utterly uncontrolled, it accounts for the interaction non-perturbatively, and it would not implausible for it to be at least qualitatively similar to the exact answer. The arguments against the possibility of turning bosonization into a controlled approximation do not rule this scenario out.

\section{Conclusion}
Patch bosonization of Fermi surfaces defines a solvable model for non-Fermi liquids. These non-Fermi liquids are rather similar to Fermi liquids in that the basic degrees of freedom are essentially phase-space densities. The mechanism for the breakdown of the Fermi-liquid in these models is rather analogous to Luttinger liquids: the elementary excitation is an electron-hole pair so strongly correlated that it cannot be broken into single fermion quasiparticles.  Unfortunately, it has been shown that these models are \textit{too} similar to Fermi liquids to be directly relevant to realistic non-Fermi liquids, where the semiclassical approximation is invalid and where any kind of phase-space description becomes explicitly nonlocal.

In my view, the essential challenge presented by non-Fermi liquids is that they are just too different from free particles to be even qualitatively understood in terms of free particles. Bosonization is just one variant of the language of free particles, as are the other synthetic limits that have been considered in the literature (limits still solved in terms of the fields of the free theory and which are explicitly devised to contain the effect of interactions to simple Feynman diagrams). I suspect understanding non-Fermi liquids will require a \textit{fundamentally} strongly interacting language, presumably in terms of variables different from those describing any kind of free excitations like electrons or electron-hole pairs. Perhaps the fact that we don't have such a language is the reason why there has been no success in reconciling any of the theoretical boson-fermion models with experimentally observed ``strange metal'' behavior.

\section{Acknowledgments}
I thank Tom Banks for valuable feedback and discussions.

\newpage
\printbibliography

\end{document}